\documentclass[twocolumn]{aastex631}

\usepackage{amsmath, xspace}
\usepackage{amssymb}
\usepackage{pifont}
\usepackage{makecell}

\defcitealias{ginolfi20}{G20}

\newcommand{\kms}{\ensuremath{\rm{km\,s}^{-1}}\xspace}
\newcommand{\sigsfr}{\ensuremath{\Sigma_{\rm{SFR}}}\xspace}
\newcommand{\cii}{[C{\scriptsize II}]\xspace}
\newcommand{\nii}{[N{\scriptsize II}]\xspace}

\newcommand{\oiii}{[O{\scriptsize III}]\xspace}
\newcommand{\etal}{et~al.\xspace}
\newcommand{\Halpha}{H\ensuremath{\alpha}\xspace}
\newcommand{\Hbeta}{H\ensuremath{\beta}\xspace}
\newcommand{\vout}{\ensuremath{v_{\rm out}}\xspace}
\newcommand{\moutdot}{\ensuremath{\dot{M}_{\rm out}}\xspace}
\newcommand{\deltabic}{\ensuremath{\Delta{\rm BIC}}\xspace}

\usepackage{nicematrix}

\BeforeBegin{NiceTabular}{\let\begin\BeginEnvironment\let\end\EndEnvironment}
\BeforeBegin{NiceArray}{\let\begin\BeginEnvironment}
\BeforeBegin{NiceMatrix}{\let\begin\BeginEnvironment}

\begin{document}

\title{The ALMA-CRISTAL survey: weak evidence for star-formation driven outflows in $z$\,$\sim$\,5 main-sequence galaxies}

\correspondingauthor{Jack Birkin}
\email{jbirkin@tamu.edu}

\author[0000-0002-3272-7568]{Jack E. Birkin}
\affiliation{Department of Physics and Astronomy and George P. and Cynthia Woods Mitchell Institute for Fundamental Physics and Astronomy, Texas A\&M University, 4242 TAMU, College Station, TX 77843-4242, US}

\author[0000-0003-3256-5615]{Justin~S.~Spilker}
\affiliation{Department of Physics and Astronomy and George P. and Cynthia Woods Mitchell Institute for Fundamental Physics and Astronomy, Texas A\&M University, 4242 TAMU, College Station, TX 77843-4242, US}

\author[0000-0002-2775-0595]{Rodrigo Herrera-Camus}
\affiliation{Departamento de Astronom\'ia, Universidad de Concepci\'on, Barrio Universitario, Concepci\'on, Chile}

\author[0000-0003-0645-5260]{Rebecca L. Davies}
\affiliation{Centre for Astrophysics and Supercomputing, Swinburne University of Technology, Hawthorn, VIC 3122, Australia}
\affiliation{ARC Centre of Excellence for All Sky Astrophysics in 3 Dimensions (ASTRO 3D)}

\author[0000-0001-7457-4371]{Lilian L. Lee}
\affiliation{Max-Planck-Institute f\"ur extratarrestrische Physik, Giessenbachstrasse 1, D-85748 Garching, Germany}

\author[0000-0002-6290-3198]{Manuel Aravena}
\affiliation{Instituto de Estudios Astrof\'{\i}sicos, Facultad de Ingenier\'{\i}a y Ciencias, Universidad Diego Portales, Av Ej\'ercito 441, Santiago, Chile}
\affiliation{Millenium Nucleus for Galaxies (MINGAL)}

\author[0000-0002-9508-3667]{Roberto J. Assef}
\affiliation{Instituto de Estudios Astrof\'{\i}sicos, Facultad de Ingenier\'{\i}a y Ciencias, Universidad Diego Portales, Av Ej\'ercito 441, Santiago, Chile}

\author[0000-0003-0057-8892]{Loreto Barcos-Mu\~{n}oz}
\affiliation{National Radio Astronomy Observatory, 520 Edgemont Road, Charlottesville, VA 22903, USA}
\affiliation{Department of Astronomy, University of Virginia, 530 McCormick Road, Charlottesville, VA 22903, USA}

\author[0000-0002-5480-5686]{Alberto Bolatto}
\affiliation{Department of Astronomy, University of Maryland, College Park, MD 20742, USA}

\author[0000-0003-0699-6083]{Tanio Diaz-Santos}
\affiliation{Institute of Astrophysics, Foundation for Research and Technology-Hellas (FORTH), Heraklion, 70013, Greece}
\affiliation{School of Sciences, European University Cyprus, Diogenes street, Engomi, 1516 Nicosia, Cyprus}

\author[0000-0002-9382-9832]{Andreas L. Faisst}
\affiliation{Caltech/IPAC, 1200 E. California Blvd. Pasadena, CA 91125, USA}

\author[0000-0002-9400-7312]{Andrea Ferrara}
\affiliation{Scuola Normale Superiore, Piazza dei Cavalieri 7, 56126 Pisa, Italy}

\author[0000-0003-0645-5260]{Deanne B. Fisher}
\affiliation{Centre for Astrophysics and Supercomputing, Swinburne University of Technology, Hawthorn, VIC 3122, Australia}
\affiliation{ARC Centre of Excellence for All Sky Astrophysics in 3 Dimensions (ASTRO 3D)}

\author[0000-0003-3926-1411]{Jorge González-López}
\affiliation{Instituto de Astrof\'isica, Facultad de F\'isica, Pontiﬁcia Universidad Cat\'olica de Chile, Santiago 7820436, Chile}
\affiliation{Las Campanas Observatory, Carnegie Institution of Washington,  Ra\'ul Bitr\'an 1200, La Serena, Chile}

\author[0000-0002-2634-9169]{Ryota Ikeda}
\affiliation{Department of Astronomy, School of Science, SOKENDAI (The Graduate University for Advanced Studies), 2-21-1 Osawa, Mitaka, Tokyo 181-8588, Japan}
\affiliation{National Astronomical Observatory of Japan, 2-21-1 Osawa, Mitaka, Tokyo 181-8588, Japan}

\author{Kirsten Knudsen}
\affiliation{Department of Space, Earth and Environment, Chalmers University of Technology, SE-412 96 Gothenburg, Sweden}

\author[0000-0002-8184-5229]{Juno Li}
\affiliation{International Centre for Radio Astronomy Research (ICRAR), The University of Western Australia, M468, 35 Stirling Highway, Crawley, WA 6009, Australia}

\author{Yuan Li}
\affiliation{Department of Physics and Astronomy and George P. and Cynthia Woods Mitchell Institute for Fundamental Physics and Astronomy, Texas A\&M University, 4242 TAMU, College Station, TX 77843-4242, US}

\author[0000-0001-9419-6355]{Ilse de Looze}
\affiliation{Sterrenkundig Observatorium, Ghent University, Krijgslaan 281 S9, B-9000 Ghent, Belgium}

\author[0000-0003-0291-9582]{Dieter Lutz}
\affiliation{Max-Planck-Institut f\"ur Extraterrestrische Physik (MPE), Giessenbachstr. 1, D-85748 Garching, Germany}

\author[0000-0001-7300-9450]{Ikki Mitsuhashi}
\affiliation{Department of Astronomy, The University of Tokyo, 7-3-1 Hongo, Bunkyo, Tokyo, 113-0033, Japan}
\affiliation{National Astronomical Observatory of Japan, 2-21-1 Osawa, Mitaka, Tokyo 181-8588, Japan}

\author[0000-0001-8598-064X]{Ana Posses}
\affiliation{Department of Physics and Astronomy and George P. and Cynthia Woods Mitchell Institute for Fundamental Physics and Astronomy, Texas A\&M University, 4242 TAMU, College Station, TX 77843-4242, US}

\author[0000-0003-1682-1148]{Monica Rela\~{n}o}
\affiliation{Departamento F\'{i}sica Te\'{o}rica y del Cosmos, Universidad de Granada, E-18071 Granada, Spain}
\affiliation{Instituto Univers\'{i}tario Carlos I de F\'{i}sica Te\'{o}rica y Computacional, Universidad de Granada, E-18071 Granada, Spain}

\author[0000-0001-6629-0379]{Manuel Solimano}
\affiliation{Instituto de Estudios Astrof\'{\i}sicos, Facultad de Ingenier\'{\i}a y Ciencias, Universidad Diego Portales, Av Ej\'ercito 441, Santiago, Chile}

\author[0000-0001-9728-8909]{Ken-ichi Tadaki}
\affiliation{Faculty of Engineering, Hokkai-Gakuen University, Toyohira-ku, Sapporo 062-8605, Japan}
\affiliation{National Astronomical Observatory of Japan, 2-21-1 Osawa, Mitaka, Tokyo 181-8588, Japan}

\author[0000-0002-5877-379X]{Vicente Villanueva}
\affiliation{Departamento de Astronom\'ia, Universidad de Concepci\'on, Barrio Universitario, Concepci\'on, Chile}

%
%
\begin{abstract}
There is a broad consensus from theory that stellar feedback in galaxies at high redshifts is essential to their evolution, alongside conflicting evidence in the observational literature about its prevalence and efficacy. To this end, we utilize deep, high-resolution [C{\sc ii}] emission line data taken as part of the [C{\sc ii}] resolved ISM in star-forming galaxies with ALMA (CRISTAL) survey. Excluding sources with kinematic evidence for gravitational interactions, we perform a rigorous stacking analysis of the remaining 15 galaxies to search for broad emission features that are too weak to detect in the individual spectra, finding only weak evidence that a broad component is needed to explain the composite spectrum. Additionally, such evidence is mostly driven by CRISTAL-02, which is already known to exhibit strong outflows in multiple ISM phases. Interpreting modest residuals in the stack at $v$\,$\sim$\,300\,km\,s$^{-1}$ as an outflow, we derive a mass outflow rate of \moutdot\,$=$\,26\,$\pm$\,11\,M$_\odot$\,yr$^{-1}$ and a cold outflow mass-loading factor of $\eta_m$\,$=$\,0.49\,$\pm$\,0.20. This result holds for the subsample with the highest star-formation rate surface density (\sigsfr\,$>$\,1.93\,M$_\odot$\,yr$^{-1}$\,kpc$^{-2}$) but no such broad component is present in the composite of the lower-star-formation rate density subsample. Our results imply that the process of star-formation-driven feedback may already be in place in typical galaxies at $z$\,$=$\,5, but on average not strong enough to completely quench ongoing star formation.
\end{abstract}

\keywords{Galaxy evolution (594), Galaxy kinematics (602), High-redshift galaxies (734), Stellar feedback (1602)}



%
%
\section{Introduction}
\label{sec:intro}

There is a general consensus that stellar feedback is an important component in galaxy formation models, and that such feedback (alongside AGN feedback) plays a key role in regulating the evolution of galaxies and producing the conditions we observe \citep[e.g.][]{dekel1986}. Massive stars inject enormous levels of energy into the interstellar medium (ISM) through supernovae and stellar winds, heating the gas and potentially expelling it from the galaxy altogether, with the possibility of later recycling it to drive further star formation \citep[e.g.][]{oppenheimer10,henriques13,angles-alcazar17}. Such feedback primarily prevents runaway star formation, regulates the metallicity of the galaxy \citep[e.g.][]{tremonti04,dalcanton07,finlator08,lilly13,torrey19}, and enables quenching. This regulation manifests in the observed mass-metallicity relation \citep[MZR; e.g.][]{tremonti04,curti20a} and fundamental metallicity relation \citep[FMR; e.g.][]{lequeux1978,ellison08,mannucci10}. Massive galaxy-scale outflows are one of the most dramatic consequences of feedback, and therefore have a huge influence on the evolution of galaxies. This process should therefore be understood and accurately implemented in theoretical models.

Observationally, outflow signatures have been detected in multiple phases of the ISM, from cool molecular and cold dust components all the way up to hot X-ray-emitting gas \citep[for a review, see e.g.][]{veilleux20}. Therefore, there are various observational indicators that can be used to search for outflow signatures. Efforts to detect outflows in galaxies and characterize their effects on the evolution of the host have been ongoing for decades. However, at high redshifts these observational indicators become fainter and much more difficult to detect. Most studies at $z$\,$\sim$\,5 and beyond have therefore focused on extreme galaxies and/or quasars \citep[e.g.][]{maiolino12,cicone15,bischetti19}. In less extreme galaxies, statistical conclusions about the nature and impact of outflows are fewer and further between \citep[although see e.g.][for studies at intermediate redshifts]{davies19,forster-schreiber19,leung19,swinbank19,weldon24}.

At $z$\,$\sim$\,4 and beyond, the emission line of singly ionized carbon at 158\,$\mu$m, \cii is one of the strongest lines that can be observed, originating primarily from neutral regions but with a non-negligible contribution from ionized gas \citep{croxall17,diaz-santos17,cormier19}. Importantly for studies at $z$\,$\sim$\,4--6, this bright line is redshifted into an atmospheric transmission window accessible from the ground by interferometers such as the Atacama Large Millimeter/submillimeter Array (ALMA). As a consequence, the \cii emission line has been exploited frequently to study outflows in sources at these epochs, most prominently high-$z$ QSOs \citep[e.g.][]{maiolino12,cicone15,bischetti19,novak20} but increasingly for star-forming galaxies in recent years \citep[e.g.][]{gallerani18,sugahara19,ginolfi20,spilker20,herrera-camus21,kade23}. On the contrary, the reliability of the \cii line as a tracer of outflows was questioned by \cite{spilker20}, who found no broad \cii emission in sources displaying clear OH outflows.

A common thread between high-redshift [C{\sc ii}] studies of outflows in QSOs and star-forming galaxies is the lack of a general consensus on their presence, absence, or ubiquity. Several studies of individual QSOs have found evidence for broad wings, such as \cite{tripodi22} whose study of the $z$\,$\sim$\,6 QSO SDSS J2310+1855 was also supported by H$_2$O emission line observations, and \cite{maiolino12} who reported an outflow in SDSS J1148+5251 at $z$\,$=$\,6.4189 \citep[later followed up and corroborated by][]{cicone15}. This result was later challenged by \cite{meyer22} using new NOEMA data. However, the QSOs studied in those works are extremely bright, making the detection of weaker broad line components in the line easier than for fainter star-forming galaxies \citep[although see][for an example of an outflow detected in an individual star-forming galaxy at $z$\,$\sim$\,5.5]{herrera-camus21}.

As a result of this limitation, authors have resorted to stacking techniques to improve the signal-to-noise (S/N) of their data, in an attempt to detect broad components in fainter sources \citep[e.g.][]{decarli18,bischetti19,davies19,stanley19,jolly20,novak20}. This method amplifies weak features in the emission but sacrifices information about individual sources, providing only average properties of the sample as a whole. An additional challenge arises from the intrinsic diversity among galaxies, including differences in size, brightness, and kinematics. These variations must be carefully accounted for when combining spectra to avoid skewing the interpretation toward galaxies with more extreme characteristics.

For QSOs there is a considerable amount of literature on \cii stacking analyses, but with differing conclusions. \cite{bischetti19} stacked a sample of 48 $z$\,$\sim$\,4.5--7.1 QSOs and found evidence of outflows with velocities greater than 1000\,km\,s$^{-1}$, corresponding to an atomic gas mass outflow rate of $\dot{M}$\,$\sim$\,100--200\,M$_\odot$\,yr$^{-1}$. In contrast, \cite{decarli18} stacked the spectra of 27 $z$\,$\sim$\,6 QSOs and found no such evidence. \cite{novak20} came to the same conclusion from a $uv$-plane stack of a similar sample of 27 $z$\,$\sim$\,6 QSOs (and a further 20 archival sources observed at lower resolution). As outflows are not likely to be isotropic, \cite{stanley19} tested the effects of orientation on stacking QSO spectra, after themselves finding only tentative evidence for broad emission from a sample of 26 $z$\,$\sim$\,6 QSOs.

The various studies employ subtly different techniques, making it difficult to compare and to determine which are most likely to be accurate. Nevertheless, studies of high-redshift star-forming galaxies, particularly those on the star-formation "main sequence", are heavily reliant on stacking simply because the sources are fainter.

In one of the most ambitious attempts to characterize the typical galaxy population at $z$\,$\sim$\,5, the ALMA Large Program to INvestigate \cii at Early times \citep[ALPINE; PI Le F\`{e}vre;][]{lefevre20,bethermin20,faisst20} detected 75 galaxies in \cii emission, which was later exploited by \cite{ginolfi20} \citepalias[hereafter][]{ginolfi20} to examine outflows at this epoch. \citetalias{ginolfi20} carried out a stacking analysis using 50 galaxies from the ALPINE sample, excluding 25 sources morpho-kinematically identified as mergers. They tested three different techniques: stacking the residuals from single-Gaussian fits to the data, stacking the [C{\sc ii}] spectra, and stacking the cubes. They also binned their sample by star-formation rate, finding only the composite of galaxies with higher star-formation rates to display broad emission, and interpreted this as evidence for star-formation driven outflows. This followed a similar analysis by \cite{gallerani18} of nine $z$\,$\sim$\,5.5 galaxies which showed flux excess in the combined residuals extended over $\sim$\,1000\,km\,s$^{-1}$.

The work of the ALPINE survey has been naturally extended through the \cii Resolved ISM in Star-forming Galaxies with ALMA (CRISTAL; Herrera-Camus \etal in prep.) Large Program that has provided higher spatial resolution (by over a factor of 2 on average) \cii data of a subsample of ALPINE galaxies, in addition to a supplementary six literature galaxies. The latter includes some galaxies from the samples studied by \cite{gallerani18} and \cite{sugahara19}. The deeper data should make it easier to detect outflows, if they exist, and the higher angular resolution will enable us to more robustly identify mergers, and to spatially resolve any outflows that are present on smaller scales.

In this work, we present a stacking analysis of the CRISTAL data, with the aim of building upon the work of \citetalias{ginolfi20}. The outline of this paper is as follows: in \S\ref{sec:observations} we describe the observations carried out along with our data reduction methods, and in \S\ref{sec:analysis} we describe our analysis of the reduced data. In \S\ref{sec:results} we present the results and in \S\ref{sec:discussion} we discuss their implications. In \S\ref{sec:conclusions} we summarize our findings. Throughout this paper we adopt a flat $\Lambda$CDM cosmology with $\Omega_\mathrm{M}$\,$=$\,0.3, $\Omega_\Lambda$\,$=$\,0.7 and $H_0$\,$=$\,70\,km\,s$^{-1}$\,Mpc$^{-1}$.

%
%
\section{Observations and data reduction}
\label{sec:observations}

\begin{table*}[]
    \centering
    \setlength\extrarowheight{1.5pt}
    \begin{NiceTabular}{c|c|c|c|c|c|c} \hline \hline
    CRISTAL ID & Alternate ID & \thead{R.A. \\ (J2000)} & \thead{Dec \\ (J2000)} & $z_{\rm [CII]}$ & \thead{log$_{10}$(SFR) \\ {[}M$_\odot$\,yr$^{-1}${]}} & \thead{log$_{10}$($M_{\rm \ast}$) \\ {[}M$_\odot${]}} \\ \hline \hline
    \multicolumn{7}{c}{Sources included in G20 composite} \\ \hline
    CRISTAL-08 & vuds\_efdcs\_530029038 & 03:32:19.03 & $-$27:52:37.6 & 4.430 & 1.88\,$\pm$\,0.23 & 9.85\,$\pm$\,0.36 \\ 
    CRISTAL-09a & DEIMOS\_COSMOS\_519281 & 09:59:00.89 & $+$02:05:27.6 & 5.575 & 1.51\,$\pm$\,0.32 & 9.84\,$\pm$\,0.39 \\ 
    CRISTAL-11 & DEIMOS\_COSMOS\_630594 & 10:00:32.59 & $+$02:15:28.4 & 4.439 & 1.57\,$\pm$\,0.31 & 9.68\,$\pm$\,0.33 \\ 
    CRISTAL-12 & CANDELS\_GOODSS\_21 & 03:32:11.95 & $-$27:41:57.5 & 5.572 & 0.98\,$\pm$\,0.40 & 9.30\,$\pm$\,0.47 \\ 
    CRISTAL-19 & DEIMOS\_COSMOS\_494763 & 10:00:05.11 & $+$02:03:12.2 & 5.233 & 1.45\,$\pm$\,0.36 & 9.51\,$\pm$\,0.36 \\ 
    \hline
    \multicolumn{7}{c}{Sources not included in G20 composite} \\ \hline
    CRISTAL-02 & DEIMOS\_COSMOS\_848185 & 10:00:21.50 & $+$02:35:11.0 & 5.294 & 2.25\,$\pm$\,0.42 & 10.30\,$\pm$\,0.28 \\ 
    CRISTAL-03 & DEIMOS\_COSMOS\_536534 & 09:59:53.26 & $+$02:07:05.5 & 5.689 & 1.79\,$\pm$\,0.31 & 10.40\,$\pm$\,0.29 \\ 
    CRISTAL-06b & -- & 10:01:01.00 & $+$01:48:34.9 & 4.562 & 1.07\,$\pm$\,0.33 & 9.19\,$\pm$\,0.46 \\ 
    CRISTAL-07a & DEIMOS\_COSMOS\_873321 & 10:00:04.06 & $+$02:37:35.8 & 5.154 & 1.89\,$\pm$\,0.26 & 10.00\,$\pm$\,0.33 \\ 
    CRISTAL-09b & -- & 09:59:00.78 & $+$02:05:26.9 & 5.577 & --\,$\pm$\,-- & --\,$\pm$\,-- \\ 
    CRISTAL-10a & DEIMOS\_COSMOS\_417567 & 10:02:04.13 & $+$01:55:44.4 & 5.671 & 1.86\,$\pm$\,0.20 & 9.99\,$\pm$\,0.31 \\ 
    CRISTAL-15 & vuds\_cosmos\_5101244930 & 10:00:47.66 & $+$02:18:02.2 & 4.580 & 1.44\,$\pm$\,0.24 & 9.69\,$\pm$\,0.33 \\ 
    CRISTAL-20 & HZ4 & 09:58:28.52 & $+$02:03:06.7 & 5.545 & 1.82\,$\pm$\,0.28 & 10.11\,$\pm$\,0.35 \\ 
    CRISTAL-23b & -- & 10:01:54.97 & $+$02:32:31.5 & 4.562 & 1.72\,$\pm$\,1.33 & 10.46\,$\pm$\,0.24 \\ 
    CRISTAL-23c & -- & 10:01:54.68 & $+$02:32:31.4 & 4.565 & --\,$\pm$\,-- & --\,$\pm$\,-- \\ 
    \hline \hline\end{NiceTabular}
    \caption{Table of basic properties for the 15 [C{\sc ii}]-detected galaxies used in this work. Stellar masses and star-formation rates are estimated from spectral energy distribution fitting with CIGALE, taken from \protect\cite{mitsuhashi24}. We divide the table into those included in, and excluded from the \protect\citetalias{ginolfi20} composite. CRISTAL-09b and CRISTAL-23c do not have available SED fits, therefore for the remainder of the analysis we adopt the median $M_\ast$ and SFR of the rest of the sample in these two cases.}
    \label{tab:sample}
\end{table*}

%
%
\subsection{The CRISTAL survey}
\label{sec:cristal}

\begin{figure}
\includegraphics[width=\linewidth]{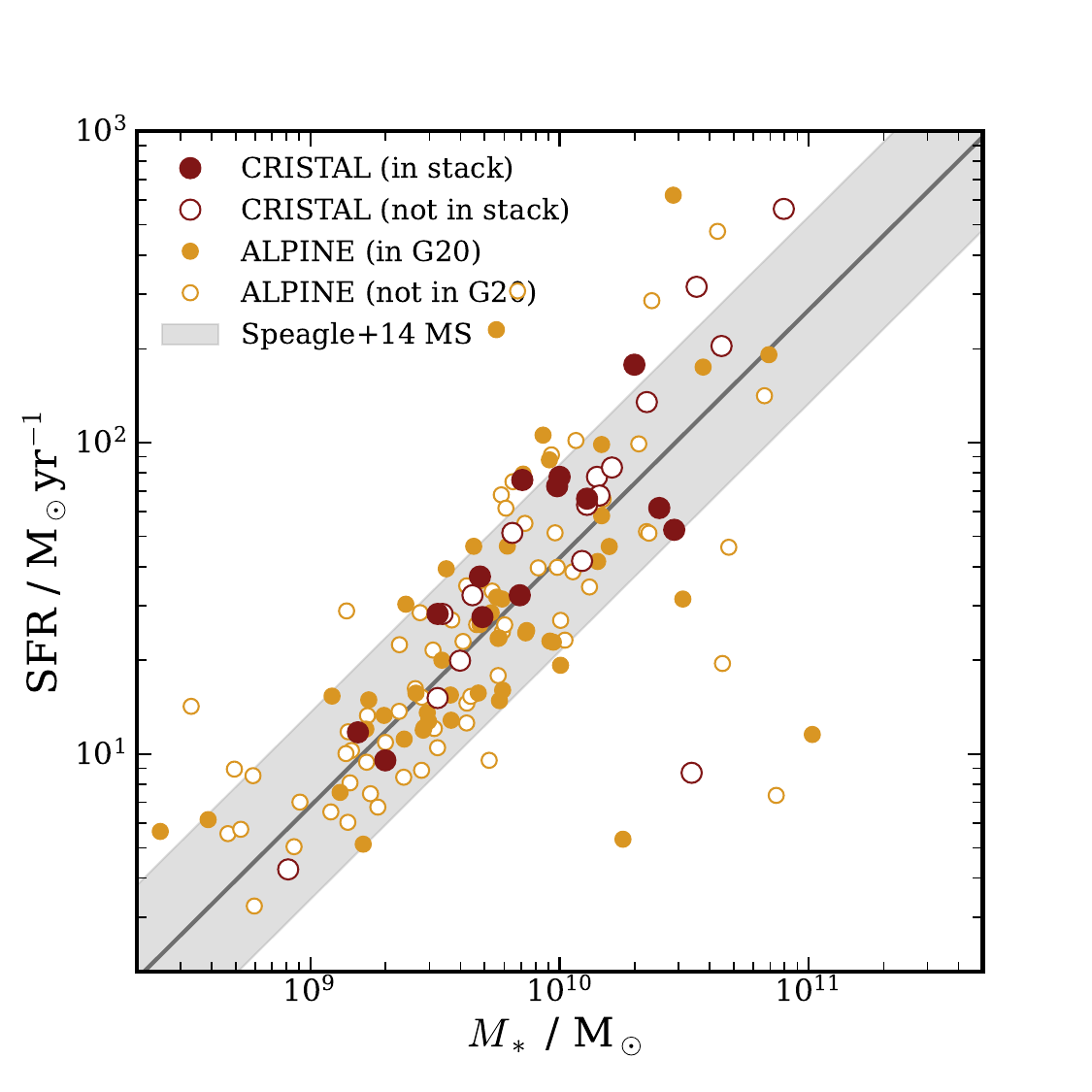}
\caption{Star-formation rate (SFR) as a function of stellar mass for the CRISTAL sample (red) along with its parent sample, ALPINE (yellow). The subset of ALPINE galaxies used in the \citetalias{ginolfi20} stack are shown as filled circles. The ALPINE sample was selected to represent sources that are close to the star-forming main sequence (shown for $z$\,$=$\,5 in gray for the \protect\cite{speagle14} prescription).}
\label{fig:main_sequence}
\end{figure}

We utilize Band 7 data from CRISTAL, an ALMA Cycle-8 Large Program (project ID 2021.1.00280.L; PI: R. Herrera-Camus) which is designed to provide spatially resolved [C{\sc ii}] and 158\,$\mu$m dust continuum maps of main sequence galaxies at $z$\,$\sim$\,5. The CRISTAL sample consists of 25 main-sequence galaxies at $z$\,$\sim$\,4.4--5.7 which were selected from the ALPINE sample (project ID 2017.1.00428.L; PI: O. Le F\`{e}vre) to have existing rest-frame UV and/or optical Hubble Space Telescope (HST) detections, stellar masses greater than 10$^{9.5}$\,M$_\odot$, and specific star-formation rates differing from that of the $z$\,$=$\,5 main sequence by no more than a factor of 3. Fig.~\ref{fig:main_sequence} shows both the ALPINE and CRISTAL samples in the stellar mass-SFR plane, with the sources used in \citetalias{ginolfi20} and this work highlighted respectively. We note that CRISTAL-09b and CRISTAL-23c do not have available SED fits and are therefore not shown in Fig.~\ref{fig:main_sequence}. To account for this in our analysis we assign them the median $M_\ast$ and SFR of the remaining 13 galaxies.

The CRISTAL sample is comprised of 19 ALPINE sources \citep{lefevre20} along with a further six galaxies in the COSMOS field (HZ4, HZ7, HZ10, DC818760, DC873756, VC8326), that have ALMA data with similar spatial resolutions and sensitivities as CRISTAL (2018.1.01359.S and 2019.1.01075.S; PI: Manuel Aravena, 2018.1.01605.S; PI: Rodrigo Herrera-Camus, 2019.1.00226.S; PI: Edo Ibar). In the 25 fields observed as part of this sample, a total of 37 galaxies were identified by either resolving the main target into multiple objects or serendipitously detecting other sources. Herrera-Camus et al. (in prep.) will present details of the survey design and observational setup. For this work we conservatively removed 22 sources that resembled mergers or have significantly disturbed kinematics based on the classification by Lee at al.\ (in prep.; and see \S\ref{sec:removing_mergers}), leaving a final sample size of 15 galaxies. The IDs, coordinates, redshifts, star-formation rates and stellar masses of the remaining 15 CRISTAL sources are shown in Table~\ref{tab:sample}.

%
%
\subsection{ALMA Band 7 Data Reduction}
\label{sec:data_reduction}

The CRISTAL and ALPINE archival observations were taken using ALMA Band 7, targeting the \cii line ($\nu_{\rm rest}$\,$=$\,1900.54\,GHz). To detect the targets, ALPINE observations were taken using the most compact configurations (typically C43-1/2) which achieved an average beam size of 0.85$''$\,$\times$\,1.13$''$ \citep{lefevre20}. As one of the main goals of CRISTAL is to spatially resolve the \cii emission, the new observations utilize more extended configurations (typically C43-4/5; beam sizes 0.1--0.3$''$) with the existing compact-configuration data being used to capture large-scale structures such as extended \cii line emission \citep[more details in Herrera-Camus et al., in prep; also see][]{ikeda24}.

To achieve greater depths, the CRISTAL team combined the ALPINE and CRISTAL raw observations and re-processed them consistently, imaging the visibilities using both natural and Briggs (robust $=$ 0.5) weighting. In this work we utilize the natural-weighted data to maximize our sensitivity to weak outflow features. The median sensitivity achieved is 0.22\,mJy in 20\,km\,s$^{-1}$ channels and the median synthesized beam FWHM is 0.46$''$\,$\times$\,0.40$''$ (2.9\,$\times$\,2.5\,kpc).

\begin{figure*}
\includegraphics[width=\linewidth]{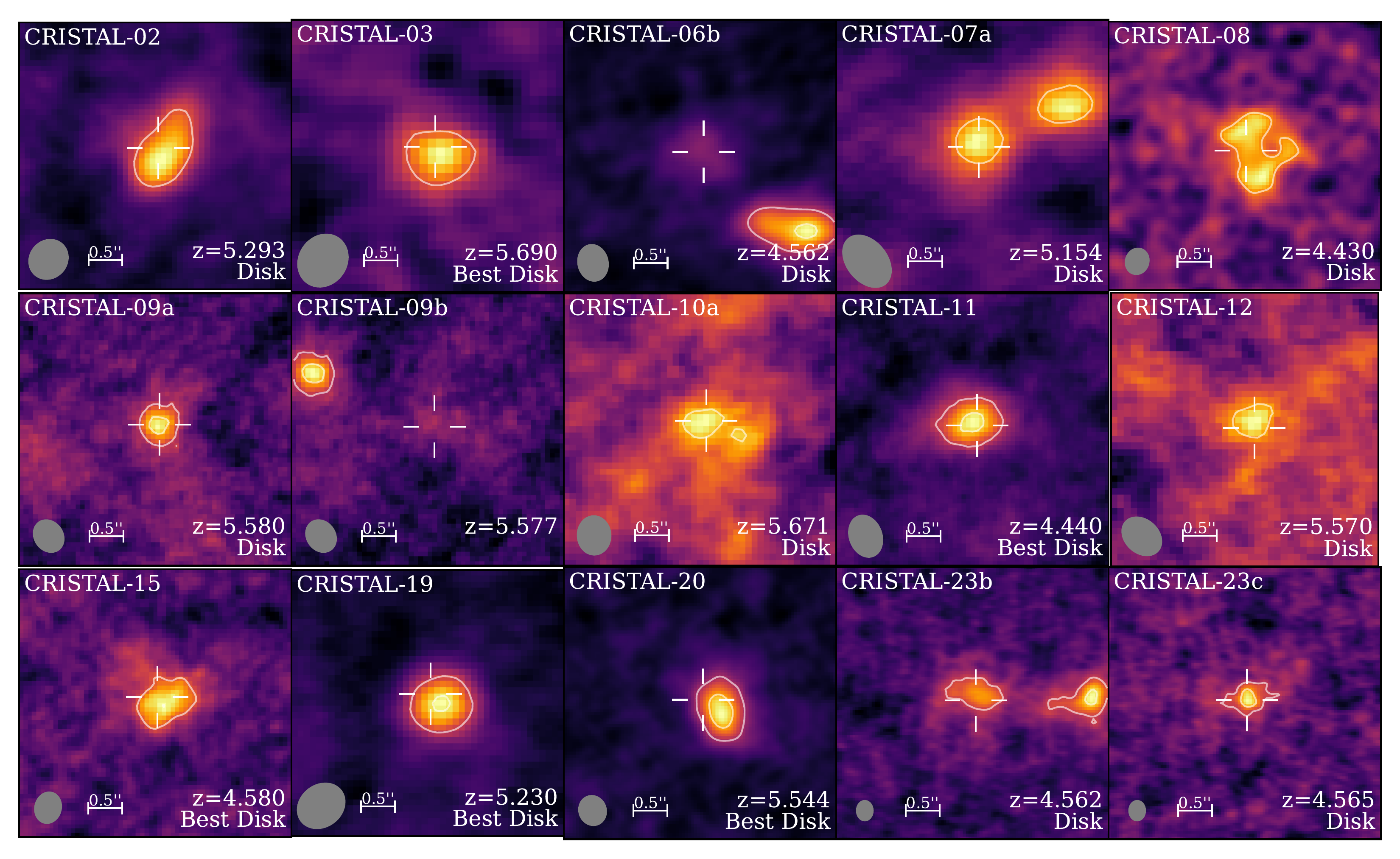}
\caption{4$''$\,$\times$\,4$''$ \cii line maps for the 15 CRISTAL galaxies that are included in the main composite. Maps are generated by collapsing the data cubes across channels spanning $\pm$\,1\,FWHM$_{\rm [CII]}$ of the \cii emission line, using the values from Herrera-Camus \etal (in prep.). The contours show \cii emission at 2\,$\sigma$ and above (increasing in 1\,$\sigma$ increments). The ALMA synthesized beam is shown in the bottom-left corner of each panel. The kinematic classifications of Lee \etal (in prep.) are shown in the bottom-right corner of each panel (CRISTAL-09b was not included in that analysis and is therefore not classified).}
\label{fig:cii_maps}
\end{figure*}

%
%
\section{Analysis}
\label{sec:analysis}

%
%
\subsection{Removing potential mergers from the sample}
\label{sec:removing_mergers}

The spectra of merging galaxies can blend multiple velocity components, from the interacting sources, disturbed gas kinematics, and the presence of tidal tails. Ultimately this results in broader spectral features which can mimic the appearance of feedback-driven outflows. Therefore, failing to account for mergers can bias the interpretation of broad emission, attributing it to outflows when it may simply reflect the dynamics of the merger. This demands a conservative approach to our analysis. \citetalias{ginolfi20} identified possible mergers in their sample using the morpho-kinematic classification of ALPINE galaxies by \cite{lefevre20}. However, as previously discussed, the angular resolution of the CRISTAL data is over twice as high as the ALPINE data, providing a new opportunity to examine which sources are likely to be mergers with higher precision.

A kinematic analysis of the full CRISTAL sample will be provided in Lee \etal (in prep.), and their classifications inform our choices of which sources to include. More specifically, they classified the CRISTAL galaxies as either disk-like, with types ``Best disk'' and ``Disk'', or non-disk. Their classification is based on a combination of kinemetry to identify deviations from orderly velocity gradients and asymmetries in the velocity/velocity dispersion maps, and spectro-astrometry to identify distinct components by utilizing the full 3-D information from the cubes.

The ``Best disk'' category comprises isolated sources with no visible neighbors or signs of perturbations; and the ``Disk'' category comprises sources that exhibit disk-like kinematics, with minor perturbations in their outskirts. Therefore, there remain sources in the ``Disk'' category that are interacting with a nearby companion, most prominently CRISTAL-07a and CRISTAL-23b (see Fig.~\ref{fig:cii_maps}). Further, separate analyses of CRISTAL-05 and CRISTAL-22 have been presented by \cite{posses24} and \cite{telikova24}, both of which appear to show complex kinematics.

Considering the above, we retain the 15 sources that are classified as ``Best disk'' and ``Disk'' sources from Lee \etal (in prep.). Based on the analysis by Lee \etal (in prep.) we do not expect interacting sources in the ``Disk'' category to contribute substantial high-velocity emission originating from these interactions to the stack, hence their inclusion. We remove the sources classified as non-disks, which lack disk-like characteristics and potentially include both major mergers and poorly resolved sources. We also conservatively remove CRISTAL-05 and CRISTAL-22 due to their complex kinematics. While this approach limits the S/N improvement achieved by stacking, it is a necessary concession to ensure that our conclusions are unbiased. The final sample is shown in Table~\ref{tab:sample}, where we separate sources that were included in/excluded from the \citetalias{ginolfi20} stack (see also Fig.~\ref{fig:main_sequence}). Interestingly, two-thirds of our sample were excluded by \citetalias{ginolfi20} on the basis of their morpho-kinematic analysis. This highlights how the differing resolution of the ALPINE and CRISTAL datasets can affect the interpretation of the same parent sample.

%
%
\subsection{Spectral extraction}
\label{sec:spectral_extraction}

For the finalized sample of 15 CRISTAL galaxies, we produce \cii line maps by collapsing the continuum-subtracted cubes within $\pm$\,1\,FWHM$_{\rm [CII]}$ of the \cii redshift (so as to ensure full coverage of the \cii emission; FWHM$_{\rm [CII]}$ measured by Herrera-Camus \etal in prep.), in a 4$''$\,$\times$\,4$''$ window centered on the source position. These are shown in Fig.~\ref{fig:cii_maps}, in which we can clearly see the morphological complexity of many of the galaxies. This makes simple circular or elliptical apertures less effective for extracting spectra than it would be for point-like sources. Therefore for each line map we generate a mask for all pixels with flux density exceeding 2\,$\sigma$, and sum all emission within the mask to extract one-dimensional (1-D) spectra for each source. To quantify the uncertainty on the flux densities, we mask channels in the 1-D spectra that are within $\pm$\,1\,FWHM$_{\rm [CII]}$ of the line centroid, and calculate the root mean square (RMS) of the remaining channels (i.e. the signal-free part of the spectrum).

A visual inspection of the spectra (see Figs.~\ref{fig:stack_all_2sig}--\ref{fig:stack_high_sfr} alongside the composites; \S\ref{sec:results}) shows that some of the line profiles are complex, with a handful appearing asymmetric. In these cases, fitting Gaussian profiles would not give a representative model of the \cii emission, and in general we prefer to derive the zeroth, first and second moments of the spectra to estimate line fluxes, centroids and linewidths, respectively \citep[e.g.][]{birkin21}. This is particularly important in deriving the velocity centroid, which must be done correctly to center each emission line before the stack. If the line centers are not aligned before stacking, this could falsely induce broad features in the composite.

We note that uncertainties on the emission centroids were not explicitly included in our analysis. These centroid uncertainties were found to be small ($<$\,20\,km\,s$^{-1}$ for the entire sample), substantially smaller than the measured line widths (median 260\,km\,s$^{-1}$). Nevertheless, to confirm that these uncertainties do not influence our results, we incorporated centroid errors into the Monte Carlo uncertainty analysis described above. We found that our conclusions remained unchanged.

As the S/N of the individual spectra are modest, median S/N\,$=$\,14\,$\pm$\,3, when deriving moments we first fit an approximate Gaussian to the line profile to establish the FWHM, and then only use channels between $\pm$\,1.5\,$\times$ the FWHM of the Gaussian to mitigate the influence of the continuum on the measured centroid \citep{birkin21}. For the 15 CRISTAL galaxies used in the stack we measure a median FWHM$_{\rm [CII]}$\,$=$\,260$^{+50}_{-80}$\,km\,s$^{-1}$, where the uncertainties reflect the 16th--84th percentile scatter in the sample. To prepare the individual spectra for stacking we shift each velocity axis by a value equal to the first moment.

%
%
\subsection{Stacking analysis}
\label{sec:stacking}
    
To combine the individual spectra we sum the flux density in each channel, weighted by the uncertainty $\sigma$ (as calculated in \S\ref{sec:spectral_extraction}) as:
\begin{equation}
    S^{\rm stack}_i = \dfrac{\Sigma^N_{k=1} S_{i,k} \times w_k}{\Sigma^N_{k=1} w_k},
    \label{eq:stack_spectra}
\end{equation}
where $S$ and $w=1/\sigma^2$ are the flux density and corresponding weight, $i$ represents the channel index, and $k$ represents the source index \citep{ginolfi20}. We sample the composite in bins of 50\,km\,s$^{-1}$, ensuring that each bin contains a reasonable number of channels from the original 20\,km\,s$^{-1}$ data. To estimate the uncertainty in each composite channel we perturb each flux density value by a random multiple of its associated error (sampled from the normal distribution) 100 times, then calculate the RMS of the resultant values.

It is not obvious what the ``correct'' method of normalizing the sample is, given the variety of properties, and therefore we attempt several methods:
\begin{enumerate}
    \item {\bf No normalization:} purely stacking the spectra as they are. This method aligns most closely with that used by \citetalias{ginolfi20}.
    \item {\bf Scaling the linewidth by the median:} it has been shown by \cite{jolly20} that stacking lines with different widths results in a non-Gaussian composite. Therefore we measure the FWHM of each line profile from the second moment, and then stretch the emission line from each source to match the median of the sample, FWHM$_{\rm med}$\,$=$\,260\,km\,s$^{-1}$, following \cite{decarli18} and \cite{novak20}. The resultant factors applied to the individual spectral axes are all in the range 0.8--2.0.
    \item {\bf Scaling the linewidth by the median and normalizing the flux density to ensure that the total integrated flux across the emission line is conserved:} expanding on method 2), but now also adjusting the flux density by the same factor as the linewidth, so that the total integrated flux is conserved. This method should give less weight to brighter sources.
    \item {\bf Scaling the linewidth to match the median and adjusting the flux density by the peak value:} similar to method 3) but we instead divide each spectrum by the peak emission flux density. This ensures that we are purely considering the {\it profile} of the emission lines. The resultant factors applied to the individual flux densities are all in the range 0.5--5.4.
\end{enumerate}
We perform tests of these four methods to ensure that we understand how they affect our conclusions, which we present in \S\ref{sec:stacking_methods}. However, for the analysis presented in the rest of this paper we mainly utilize the composites derived using method 4.

%
%
\subsection{Line fitting}
\label{sec:line_fitting}

Having produced composite spectra, we search for broad components that may indicate outflows. To do so, we fit two models to the \cii emission line, using the Markov Chain Monte Carlo (MCMC) ensemble sampler {\tt emcee} \citep{foreman-mackey13}. The first is a simple single-Gaussian model with free parameters amplitude, mean and sigma. The second is a double-Gaussian model, one narrow and one broad, each with the same three free parameters. However, in the double-Gaussian model we place a number of constraints on these parameters:
\begin{itemize}
    \item The narrow FWHM is constrained to be between 80\,km\,s$^{-1}$ and 400\,km\,s$^{-1}$ and the broad FWHM is constrained to be between 400\,km\,s$^{-1}$ and 1500\,km\,s$^{-1}$,
    \item The central wavelengths of both lines are constrained to be within $\pm$\,200\,km\,s$^{-1}$ of the systemic redshift,
    \item The narrow amplitude is constrained to be at least twice the broad amplitude \citep{carniani24},
    \item The broad FWHM is constrained to be at least 1.2\,$\times$ the narrow FWHM \citep{carniani24}.
\end{itemize}

To statistically conclude whether or not broad components are necessary to describe our stacks, we implement a Bayesian Information Criterion (BIC) test. Measured as BIC\,$=$\,$\chi^2+k\log{n}$, where $\chi^2$ is the chi-squared value of the model, $k$ is the number of parameters in the model and $n$ is the number of data points used in the fit, the BIC punishes the model if using many parameters to obtain a good fit. This allows us to quantitatively decide whether adding an additional model component to describe an outflow provides a statistically better description of the data to justify the extra parameters, based on the difference in BIC between the single- and double-Gaussian models, calculated as \deltabic\,$=$\,BIC$_{\rm single}$\,$-$\,BIC$_{\rm double}$. Although the \deltabic test does not give a statistical significance such as a p-value, the general rule of thumb is that \deltabic\,$>$\,10 represents overwhelming evidence for the model with the lowest BIC. In our analysis, we will measure outflow properties if the \deltabic\,$>$\,0, but require \deltabic\,$>$\,10 to securely reject the null hypothesis that the composite \cii emission is satisfactorily described by a single Gaussian \citep{swinbank19,avery21,concas22,weldon24}.

We note that this choice of \deltabic\,$=$\,10 is not a perfect one in the context of fitting outflows \citep{chu24}, and it was mostly made to ensure comparability with other similar studies. For full transparency we include the measured \deltabic values with our results, and it is left up to the reader to make their own conclusions about the significance of these results.

%
%
\subsection{Bootstrap/jackknife resampling tests}
\label{sec:resampling}

Previously we have aimed to remove any obvious individual sources which may mimic outflows in the sample (see \S\ref{sec:removing_mergers}). However, we still wish to test whether any of the remaining galaxies could potentially bias our conclusions. For example, CRISTAL-02 is known to drive a powerful outflow from its central regions (Davies et al.\ in prep.). To account for this, we perform a bootstrap test by resampling with replacement. For each composite, we create 100 resampled sets by randomly selecting from the included galaxies, and then apply the BIC test described in \S\ref{sec:line_fitting}. Further, we perform a separate stack of the full sample without CRISTAL-02, the individual CRISTAL galaxy with the most obvious outflow (Davies et al.\ in prep.), reducing the sample size to 14. We discuss the results of these tests in \S\ref{sec:results}.

%
%
\section{Results}
\label{sec:results}

\begin{figure*}
\includegraphics[width=\linewidth]{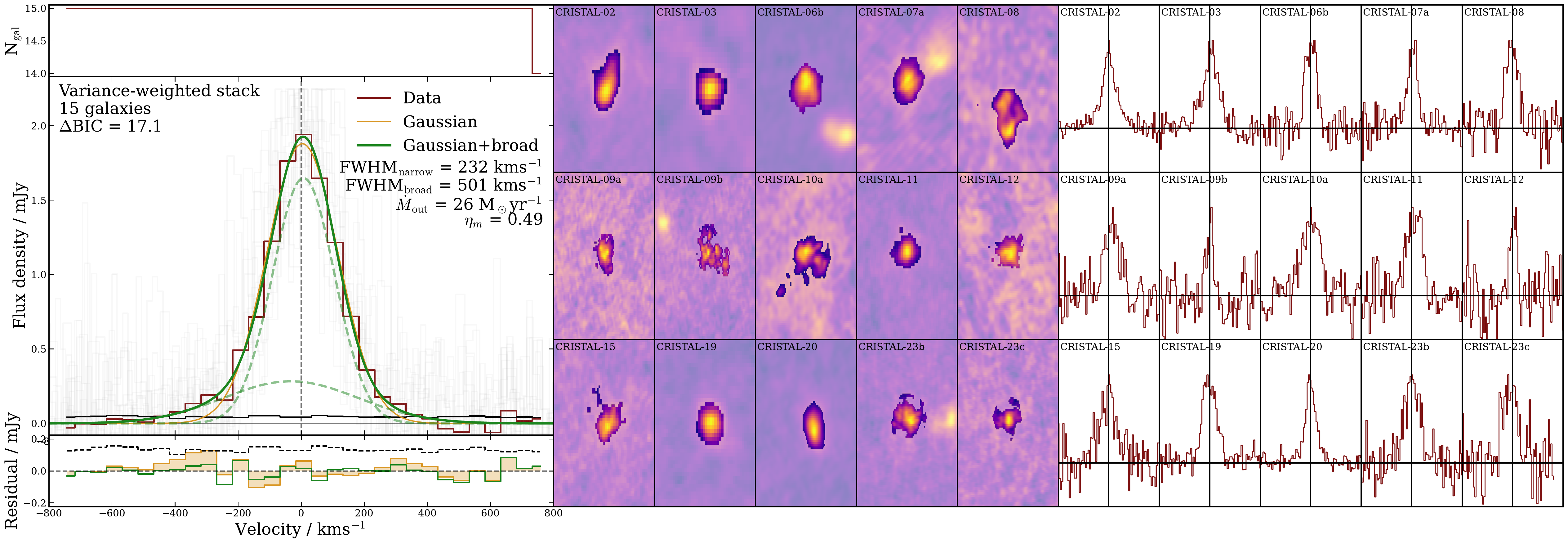}
\caption{{\it Left:} Stack of the colored regions in the \cii line maps (red) at 50\,km\,s$^{-1}$ resolution, with both single- (yellow) and double-Gaussian (green) models overlaid. We also decompose the double-Gaussian model into its individual components (green dashed). The individual spectra of the galaxies used in the stack are shown as the faded gray lines. The top panel shows the number of galaxies that contribute to each velocity bin, while the bottom panel shows the residuals from both fits. In the bottom panel, the dashed black line shows the 3-$\sigma$ uncertainty. In this case, the BIC test described in \S\ref{sec:line_fitting} indicates that a broad component is necessary to model the composite emission line.
{\it Center:} \cii emission line maps for all CRISTAL galaxies, with bold colored regions indicating the areas from which spectra were extracted.
{\it Right:} Individual spectra used in the stack, at 20\,km\,s$^{-1}$ resolution.
}
\label{fig:stack_all_2sig}
\end{figure*}

%
%
\subsection{Global stacks}
\label{sec:global_stacks}

We stack the integrated spectra of all 15 \cii-detected CRISTAL disk-like galaxies using method 4 (see \S\ref{sec:stacking}), the result of which is shown in Fig.~\ref{fig:stack_all_2sig}. The majority of the individual line profiles, also shown in Fig.~\ref{fig:stack_all_2sig}, do not reveal any obvious outflow signatures, although the double-Gaussian fit appears to better match the outer parts of the composite profile. This can be seen in the single-Gaussian residuals which show a feature at $\sim$\,$-$300\,km\,s$^{-1}$ that is not present in the double-Gaussian model. Indeed, we find \deltabic\,$=$\,17 for the composite indicating that the double-Gaussian model is preferred, which has FWHM$_{\rm narrow}$\,$=$\,232\,$\pm$\,13\,km\,s$^{-1}$ and FWHM$_{\rm broad}$\,$=$\,501\,$\pm$\,62\,km\,s$^{-1}$. Stacks of the residuals from single-Gaussian fits to each source were found to be consistent with the residuals shown in Fig.~\ref{fig:stack_all_2sig}.

\begin{figure}
    \centering
    \includegraphics[width=\linewidth]{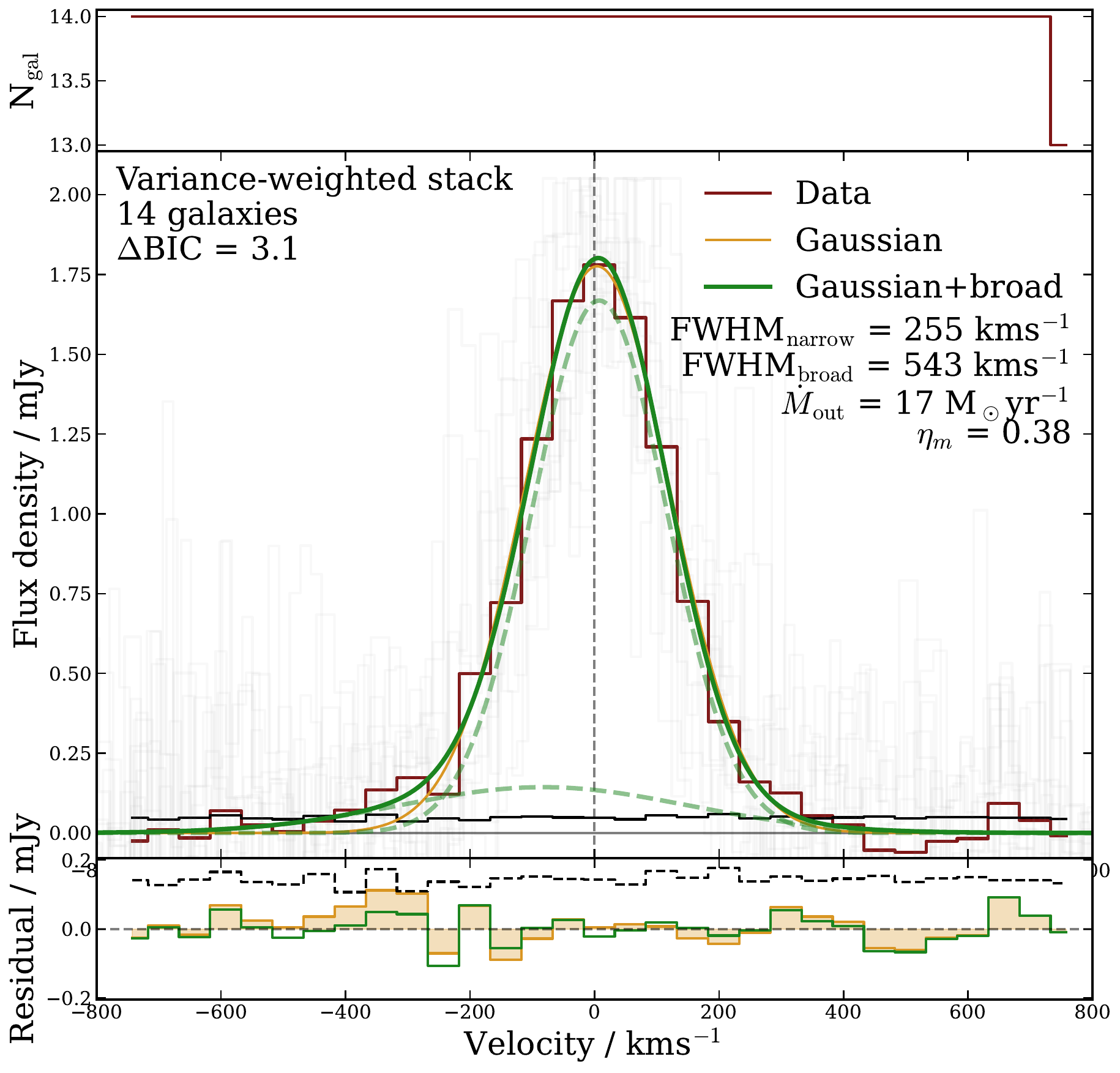}
    \caption{Stack of the same galaxies as shown in Fig.~\ref{fig:stack_all_2sig} except for CRISTAL-02 which has been removed. Given the strong outflows already detected in this source, we test to see how much it drives our results. In this case, the BIC test described in \S\ref{sec:line_fitting} indicates that a broad component is only marginally significant in the composite spectrum. This would suggest that the outflow components observed in Figs.~\ref{fig:stack_all_2sig} and \ref{fig:stack_high_sfr} are largely driven by this individual source.
    }
    \label{fig:stack_no_c02}
\end{figure}

Applying the bootstrap test described in \S\ref{sec:resampling}, we find that in 55$\%$ of cases broad emission is identified, with only 8$\%$ of all cases having \deltabic\,$>$\,10. This implies that a small number of sources, likely including CRISTAL-02, are significantly influencing the need for a broad component when modeling the stacked spectra. Indeed, when purposefully removing CRISTAL-02 from the stack we find \deltabic\,$=$\,3.1 (shown in Fig.~\ref{fig:stack_no_c02}). Given this result, and the relatively low $\deltabic$, we choose to interpret our global composite as providing only modest support for outflows.

%
%
\subsection{Resolved composites}
\label{sec:resolved_stacks}

\begin{figure*}
    \includegraphics[width=\linewidth]{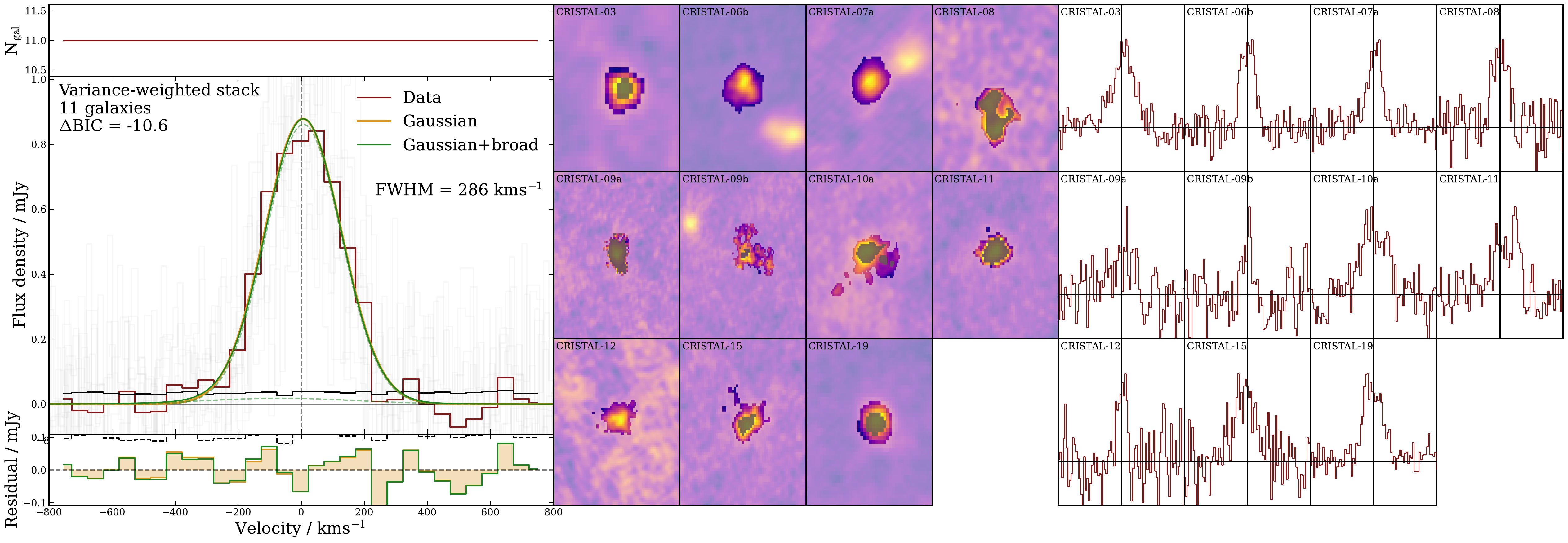}
    \caption{The same as Fig.~\ref{fig:stack_all_2sig}, but only including regions of the sample with \sigsfr\,$<$\,1.93\,M$_\odot$\,yr$^{-1}$\,kpc$^{-2}$. The composite is well-modeled by a single-Gaussian function.
    }
    \label{fig:stack_low_sfr}
\end{figure*}

\begin{figure*}
    \includegraphics[width=\linewidth]{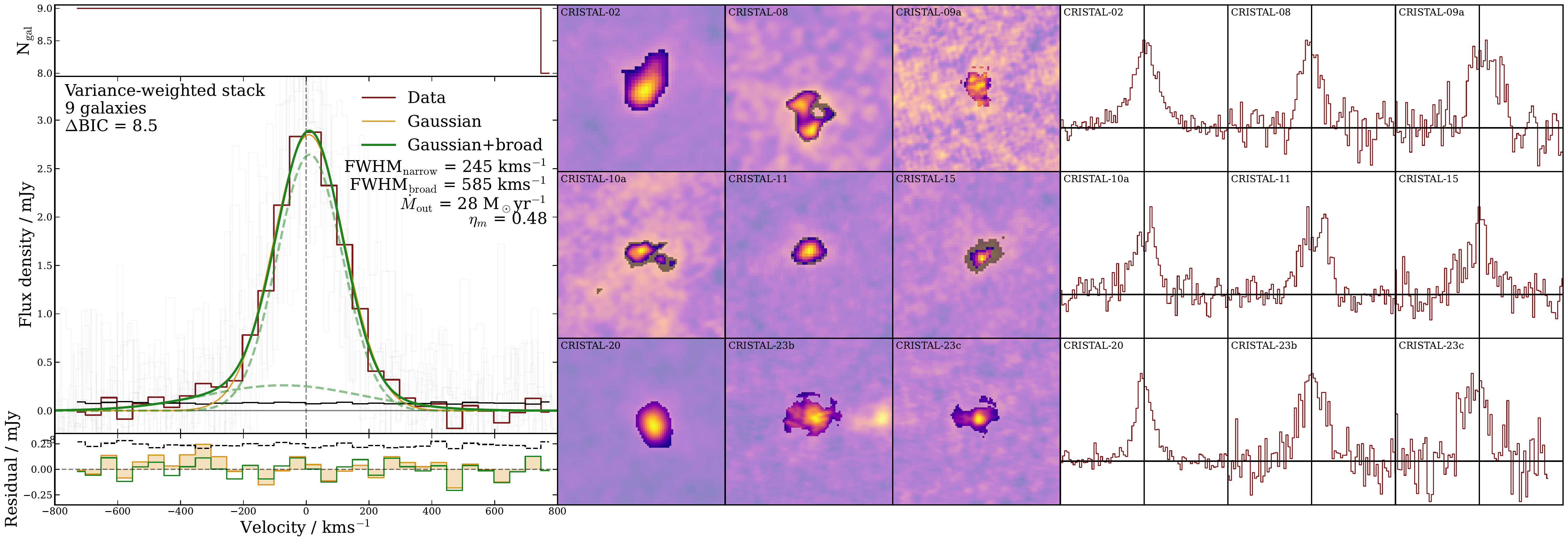}
    \caption{The same as Fig.~\ref{fig:stack_all_2sig}, but only including regions of the sample with \sigsfr\,$>$\,1.93\,M$_\odot$\,yr$^{-1}$\,kpc$^{-2}$. With \deltabic\,$=$\,9, the composite requires a broad component to fully describe the emission.
    }
    \label{fig:stack_high_sfr}
\end{figure*}

One of the key results obtained by \citetalias{ginolfi20} showed that outflows are only present in the stacks of ALPINE galaxies with higher global star-formation rates. Other authors have found a similar result \citep[see e.g.][]{forster-schreiber19,swinbank19}. This was shown by dividing the sample into low-SFR and high-SFR bins and stacking the resultant subsets. However, in reality, outflows likely occur in specific regions of galaxies rather than uniformly \citep{genzel11,newman12,davies19}.

To test these findings, we build on the \citetalias{ginolfi20} result by leveraging the improved resolution of CRISTAL to focus on stacking only {\it resolved low- and high-\sigsfr galaxy regions} across the sample. To split the sample into low- and high-\sigsfr galaxy regions we first convert all (2\,$\sigma$-masked) [C{\sc ii}] maps to \sigsfr maps using the average $L_{\rm [CII]}$--SFR trend of \cite{lagache18}:

\begin{equation}
    \log\left(\dfrac{L_{\rm [CII]}}{\rm{L}_\odot}\right) = (1.4-0.07z)\times\log\left(\dfrac{\rm SFR}{{\rm M}_\odot{\rm yr}^{-1}}\right)+7.1-0.07z,
    \label{eq:lcii-sfr}
\end{equation}

\noindent and dividing by the pixel area in kpc$^2$. At this point each spatial pixel has a corresponding value of \sigsfr. We then choose the median value of all pixels, \sigsfr\,$=$\,1.93\,M$_\odot$\,yr$^{-1}$\,kpc$^{-2}$, as the boundary between low- and high-\sigsfr galaxy regions. In this regard, low- and high-\sigsfr are meant not in the absolute sense, but in relation to the sample as a whole. Using data from the SINS/zC-SINF of $z$\,$\sim$\,2 galaxies, \cite{newman12} identified a threshold of \sigsfr\,$=$\,1\,M$_\odot$\,yr$^{-1}$\,kpc$^{-2}$ above which strong outflows occurred in their sample. As our threshold between low- and high-\sigsfr is higher than this, in theory we should see outflows in both samples, but in reality our threshold is dependent on our choice of $L_{\rm [CII]}$--SFR calibration. Given this uncertainty we prefer to interpret the results of stacking the low- and high-\sigsfr subsets in isolation to search for a trend of outflows with \sigsfr, rather than comparing to other studies.

Some galaxies naturally will have pixels that fall into both sub-samples e.g. CRISTAL-08. We impose an additional requirement that to include a galaxy in the low/high-\sigsfr subsample, the number of pixels that fall below/above the \sigsfr threshold must be greater than the number of the pixels within the synthesized beam. This is to ensure that the summed spectra have adequate S/N to be used in the stack. The regions of each \cii map that are classified as low-\sigsfr and high-\sigsfr are highlighted in Figs.~\ref{fig:stack_low_sfr} and \ref{fig:stack_high_sfr}.

Fig.~\ref{fig:stack_low_sfr} shows the composite of all regions of the sample with \sigsfr\,$<$\,1.93\,M$_\odot$\,yr$^{-1}$\,kpc$^{-2}$, which includes 11 of the 15 galaxies used in the previous stack. Contrary to Fig.~\ref{fig:stack_all_2sig}, we see no evidence for broad components in this composite, with the residuals appearing random for both models and \deltabic\,$=$\,\ensuremath{-11}. On the other hand, Fig.~\ref{fig:stack_high_sfr} shows the composite of all regions of the sample with \sigsfr\,$>$\,1.93\,M$_\odot$\,yr$^{-1}$\,kpc$^{-2}$. This sample includes 9 of the original 15 sources, including CRISTAL-02 that shows a strong broad feature in its integrated spectrum (Davies et al.\ in prep.). This composite spectrum displays a broad feature with FWHM$_{\rm narrow}$\,$=$\,245\,$\pm$\,11\,km\,s$^{-1}$ and FWHM$_{\rm broad}$\,$=$\,585\,$\pm$\,76\,km\,s$^{-1}$, broader than the full-sample stack but still consistent within the uncertainties. As with the full-sample stack, the combined residuals from single-Gaussian fits to each were consistent with the residuals from fitting a single-Gaussian model to both the low-\sigsfr and high-\sigsfr samples.
We note that it is entirely possible that this observation, namely the higher significance of broad emission in the low-\sigsfr sample, is simply due to the \cii emission being brighter in those galaxies (as a result of using \cii to trace SFR). This would make any fainter broad emission easier to detect. Despite this, the low-\sigsfr and high-\sigsfr stacks have similar S/N, 39 and 37 respectively. Therefore, both stacks should have similar ability to yield broad line detections.

We conducted a similar test by dividing each galaxy into inner and outer regions. This division was achieved by sorting flux density values into two bins {\it on a source-by-source basis}, with the fainter bin representing the outer regions and the brighter bin corresponding to the inner regions. We visually inspected the resulting masks to ensure this division was accurate. In doing so, we found statistically moderate evidence of broad emission in the inner regions (\deltabic\,$=$\,5.1), but not in the outer regions, as shown in Figs.~\ref{fig:stack_outer} and \ref{fig:stack_inner}. This suggests that outflows predominantly originate from the center of the galaxy (which are also generally the regions with the highest \sigsfr).

\begin{table*}
    \centering
    \setlength\extrarowheight{2pt}
    \begin{NiceTabular}{c|c|c|c|c|c|c|c|c}\hline\hline
    Sample & $N_{\rm gal}$ & $z_{\rm med}$ & \thead{SFR$_{\rm med}$ \\ {[}M$_\odot$\,yr$^{-1}${]}} & \thead{$M_{\rm \ast, med}$ \\ {[}10$^9$ M$_\odot${]}} & $\Delta$(BIC) & Result & \thead{FWHM$_{\rm narrow}$ \\ {[}km\,s$^{-1}${]}} & \thead{FWHM$_{\rm broad}$ \\ {[}km\,s$^{-1}${]}} \\ \hline
    \multicolumn{9}{c}{\bf Method 1: No FWHM normalization, no flux normalization} \\ \hline
    Full sample & 15 & 5.230 & 52\,$^{+24}_{-26}$ & 7\,$^{+13}_{-4}$ & 0.4 & {\it Double} & 229\,$\pm$\,12 & 543\,$\pm$\,71\\
    Low-SFR & 11 & 5.230 & 35\,$^{+40}_{-16}$ & 6\,$^{+4}_{-3}$ & -8.8 & Single & 293\,$\pm$\,14 & \\
    High-SFR & 9 & 4.580 & 59\,$^{+17}_{-26}$ & 8\,$^{+11}_{-3}$ & 8.2 & {\it Double} & 219\,$\pm$\,14 & 492\,$\pm$\,49\\
    \hline
    \multicolumn{9}{c}{\bf Method 2: FWHM normalization, no flux normalization} \\ \hline
    Full sample & 15 & 5.230 & 52\,$^{+24}_{-26}$ & 7\,$^{+13}_{-4}$ & -2.3 & Single & 268\,$\pm$\,8 & \\
    Low-SFR & 11 & 5.230 & 35\,$^{+40}_{-16}$ & 6\,$^{+4}_{-3}$ & -10.9 & Single & 297\,$\pm$\,18 & \\
    High-SFR & 9 & 4.580 & 59\,$^{+17}_{-26}$ & 8\,$^{+11}_{-3}$ & -1.3 & Single & 250\,$\pm$\,10 & \\
    \hline
    \multicolumn{9}{c}{\bf Method 3: FWHM normalization, flux conservation} \\ \hline
    Full sample & 15 & 5.230 & 52\,$^{+24}_{-26}$ & 7\,$^{+13}_{-4}$ & -7.7 & Single & 276\,$\pm$\,11 & \\
    Low-SFR & 11 & 5.230 & 35\,$^{+40}_{-16}$ & 6\,$^{+4}_{-3}$ & -11.1 & Single & 289\,$\pm$\,17 & \\
    High-SFR & 9 & 4.580 & 59\,$^{+17}_{-26}$ & 8\,$^{+11}_{-3}$ & -0.2 & Single & 256\,$\pm$\,7 & \\
    \hline
    \multicolumn{9}{c}{\bf Method 4: FWHM normalization, flux normalization} \\ \hline
    Full sample & 15 & 5.230 & 52\,$^{+24}_{-26}$ & 7\,$^{+13}_{-4}$ & 17.1 & {\bf Double} & 232\,$\pm$\,13 & 501\,$\pm$\,62\\
    Low-SFR & 11 & 5.230 & 35\,$^{+40}_{-16}$ & 6\,$^{+4}_{-3}$ & -10.6 & Single & 286\,$\pm$\,19 & \\
    High-SFR & 9 & 4.580 & 59\,$^{+17}_{-26}$ & 8\,$^{+11}_{-3}$ & 8.5 & {\it Double} & 245\,$\pm$\,11 & 585\,$\pm$\,76\\
    \hline
    \hline\end{NiceTabular}
    \caption{Table showing the properties of each stacking sub-sample: the number of galaxies in the sample, the median redshift, star-formation rate and stellar mass, the difference in BIC for single- and double-Gaussian models and corresponding conclusion on model preference, and the narrow and broad linewidths. Quoted uncertainties on the median SFR and $M_\ast$ are the 16$^{\rm th}$\,--\,84$^{\rm th}$ percentile ranges. In the result column, ``double'' is printed in italics if 0\,$<$\,\deltabic\,$<$\,10 (indicating marginal broad emission) and in bold if \deltabic\,$\ge$\,10 (indicating significant broad emission).}
    \label{tab:stack_results}
\end{table*}

%
%
\subsection{Effects of using different stacking methods}
\label{sec:stacking_methods}

As discussed in \S\ref{sec:stacking}, we chose to test four different methods of normalization when stacking the CRISTAL spectra. The main aim of this test is not to show which method is best, but to characterize how the results vary depending on the choice of normalization. As our goal is to analyze the line {\it profiles} in our sample, we have chosen to focus on presenting and analyzing the method 4 stacks, in which all lines are normalized to the same width and peak flux density. Indeed, we have already shown the results of the method 4 stacks, in which we find only moderate evidence for broad wings in the integrated spectrum of the entire sample and in the high-\sigsfr regions of the sample (predominantly the central regions). Here, we provide and compare the individual results for all four methods, which are shown in Table~\ref{tab:stack_results}.

If we stack the spectra purely as they are extracted with no changes (method 1), we see moderate evidence for broad emission in both the full sample and high-\sigsfr sample. This makes sense as not normalizing the linewidths can artificially introduce broad features \citep{jolly20}. In this stack the full sample shows FWHM$_{\rm broad}$\,$=$\,543\,$\pm$\,71\,km\,s$^{-1}$  (broader than using method 4).

After normalizing the linewidths this evidence disappears without any flux normalization (method 2). In this case we would expect any real broad emission to be more clearly distinguished. The broad emission also remains absent when peak flux is multiplied to conserve the total flux (method 3), a method which should reduce the impact of the brighter sources in the sample. The fact that method 4 gives the clearest indication of broad emission in the stack is interesting because this method gives equal weight to all sources and would therefore imply that there is a real underlying broad signal in the full sample. However, it is also very possible that the low S/N of this data is the cause of this result, and the impact of removing CRISTAL-02 from the sample reinforces this idea.

%
%
\subsection{Comparisons with G20 analysis}
\label{sec:simulations}

The detection of broad [C{\sc ii}] emission in high-redshift galaxies offers a potential signature of large-scale outflows driven by star formation. In a key study, \citetalias{ginolfi20} stacked the global [C{\sc ii}] spectra of 50 ALPINE galaxies and found that the composite spectrum is best described by the sum of a narrow Gaussian with FWHM$_{\rm narrow}$\,$=$\,233\,$\pm$\,15\,km\,s$^{-1}$ and a broad Gaussian with FWHM$_{\rm broad}$\,$=$\,531\,$\pm$\,90\,km\,s$^{-1}$. They also found even broader emission when stacking only galaxies with SFRs above the sample median.

The \citetalias{ginolfi20} analysis has informed the work presented in this paper, but there are a number of important differences between the two projects, most notably in the data quality. The CRISTAL data used in this study has a median synthesized beam FWHM of 0.46\,$\times$\,0.40$''$, over a factor of two improvement on the ALPINE data, which has a median of 0.85\,$\times$\,1.13$''$ \citep{lefevre20}. Therefore, our data probe scales of $\sim$\,3.9\,kpc at $z$\,$=$\,5, whereas for \citetalias{ginolfi20} this value is $\sim$\,9.1\,kpc. This finer resolution enables us to better resolve the internal kinematics of galaxies and mitigate the effects of beam smearing, which can artificially broaden line profiles. Indeed, five sources classified as interacting by Lee \etal (in prep.; and therefore excluded from our analysis) were included in the \citetalias{ginolfi20} composite.

To test whether the broad features reported by \citetalias{ginolfi20} would be recovered under our stacking methodology and data quality, we simulated their composite spectrum using their best-fit Gaussian parameters, added noise consistent with the RMS of our data, and stacked the resulting spectra using our method 4. For this comparison, we focus on the full-sample stack from \citetalias{ginolfi20} (their Fig.~5). Given the deeper sensitivity of the CRISTAL data (Herrera-Camus \etal in prep.), we expect to be sensitive to faint broad components in the stack.

\begin{figure}
    \centering
    \includegraphics[width=\linewidth]{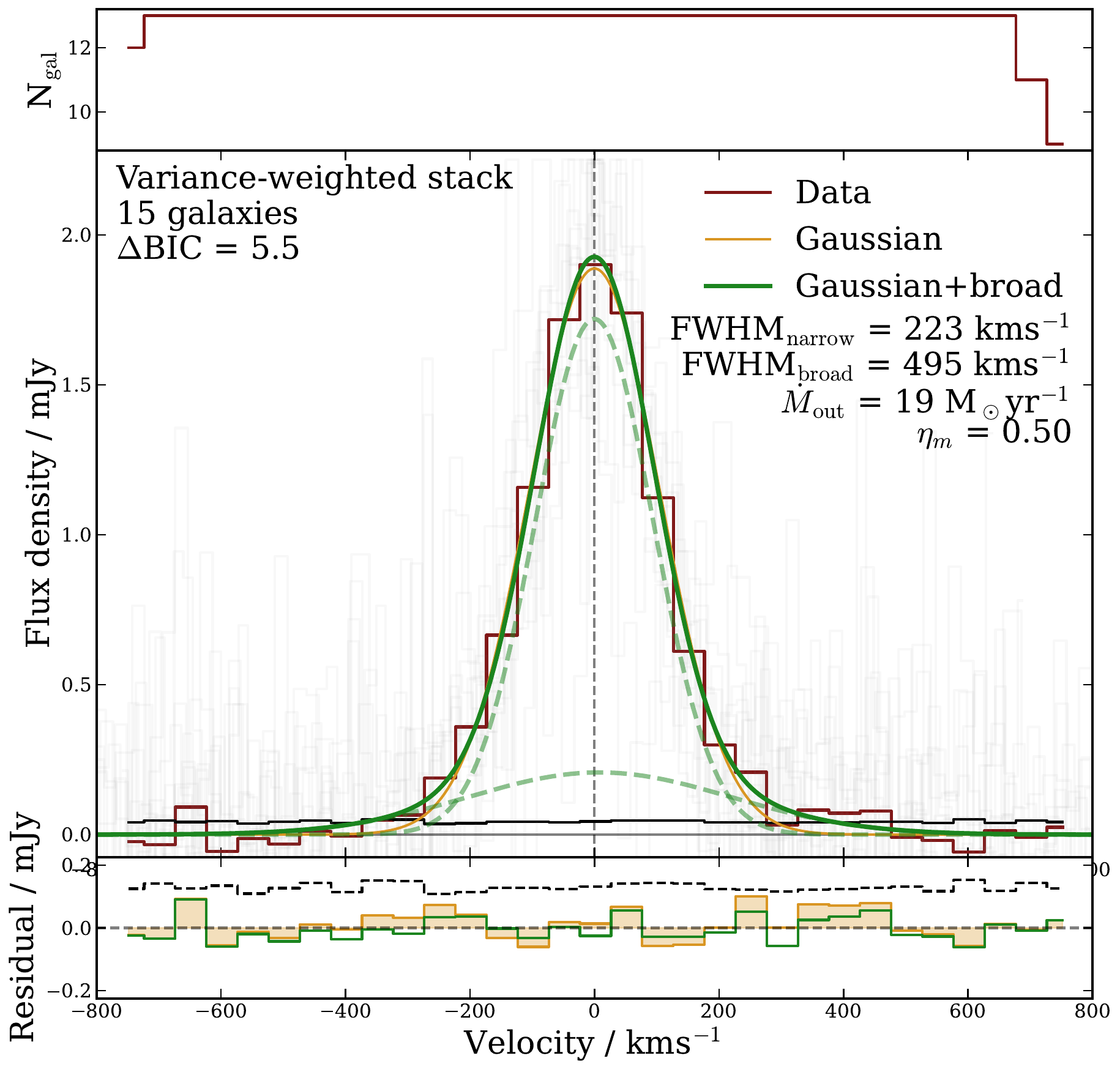}
    \caption{Simulated results obtained from inserting the \citetalias{ginolfi20} composite signal into spectra with the RMS of the CRISTAL data. The broad emission is detected with \deltabic\,$=$\,5.5, i.e. less significant than our full-sample stack.
    }
    \label{fig:g20_sims}
\end{figure}

The results of this stack are shown in Fig.~\ref{fig:g20_sims}, showing marginal broad features, slightly less significant than in our full-sample composite. We measure narrow and broad line widths of 223\,\kms and 495\,\kms respectively, lower than the input values of 233\,\kms and 531\,\kms, although this is simply because we normalized all spectra to the same linewidth as in our CRISTAL stacks. In general, Fig.~\ref{fig:g20_sims} shows that the outflow signal measured by \citetalias{ginolfi20} could be detected with the sensitivity of the CRISTAL data analyzed here. It is difficult, however, to more directly compare this result with Fig.~5 in \citetalias{ginolfi20} as they do not include \deltabic values, but in general we do not see residuals above 3-$\sigma$, whereas they do. Several differences in sample selection, resolution, and methodology may explain these residual discrepancies between the two analyses.

Firstly, the sample utilized by \citetalias{ginolfi20} is much larger than ours, 50 versus 15 (11 and 9 in the low- and high-\sigsfr stacks respectively). As a result, the S/N gain from stacking should be a factor of $\sim$\,1.8 better in the \citetalias{ginolfi20} analysis compared to ours. It is also possible that the ALPINE stacking sample includes sources with stronger outflows on average than those selected by CRISTAL. This seems unlikely if, as \citetalias{ginolfi20} suggests, the outflows are primarily driven by star formation, because the CRISTAL sources are among the highest-SFR sources of the full ALPINE sample (see Fig.~\ref{fig:main_sequence}). A further possibility is that there remain incorrectly classified mergers in the \citetalias{ginolfi20} sample. Given that these sources likely have higher SFRs than non-merging sources, this could contribute to their interpretation that outflows are star-formation driven.

Another difference is that the coarser spatial resolution of the original ALPINE data could artificially induce broad components in the spectra, potentially from mergers. The kinematic analysis of the higher-resolution CRISTAL data provided by Lee \etal (in prep.) has allowed us to more accurately exclude mergers from the sample. Unfortunately this comes at the cost of a reduced sample size, meaning that stacking provides a smaller increase in S/N. Due to the mass-based selection of CRISTAL, the proportion of interacting sources in our sample is expected to be higher than that of ALPINE. 

Additionally, the two analyses employ different methodologies. \citetalias{ginolfi20} did not normalize the linewidths in their stack (most closely resembling our method 1, which could have artificially introduced broad components into their composite \citep{jolly20}. However, we tested this method as part of our analysis (see \S\ref{sec:stacking_methods}) and found that the results did not skew significantly towards double-Gaussian when compared to our method 4.

In summary, while differences in resolution, sample selection, and stacking methodology can influence the interpretation of broad emission components, both studies point toward the tentative presence of galaxy-scale outflows in normal star-forming galaxies at $z$\,$\sim$\,5. A definitive understanding of these outflows will require large, statistically robust samples observed at CRISTAL-like resolution. Such a study is well within the capability of ALMA and would represent a critical next step in understanding feedback processes in the early universe.

%
%
\section{Discussion}
\label{sec:discussion}

We now discuss our findings from the various stacks and their implications for the presence or absence of feedback in $z$\,$\sim$\,5 main-sequence star-forming galaxies.

%
%
\subsection{Are outflows responsible for the broad emission?}
\label{sec:outflows?}

While we have found modest evidence that a broad component is required to explain the \cii emission in the CRISTAL sample, this does not preclude scenarios other than outflows. In this section we discuss other possibilities and justify interpreting the broad component as an outflow.

On the marginality of the broad detection, we address here the possible biasing of our sample selection. Given our conservative approach to rejecting potential mergers, our final sample comprises strictly disk-like sources with measured inclinations in the range 43$^\circ$--84$^\circ$ (Lee \etal in prep.), meaning that most are closer to being edge-on disks. If true, outflows perpendicular to the disk will have a relatively low velocity along the line-of-sight and therefore be more difficult to detect. This is briefly discussed further in \S\ref{sec:outflow_velocity}.

As discussed in \S\ref{sec:removing_mergers}, the spectra of merging galaxies can combine multiple velocity components due to interacting sources, disturbed gas kinematics, and tidal tails, resulting in broader spectral features that can mimic feedback-driven outflows. While mergers and interactions are known to produce such broad features, our careful approach ensures these effects are minimized. If indeed there are any mergers remaining in the sample after the selection, they must be close mergers ($<$\,0.4$''$) with a small offset between velocity components ($\sim$\,100\,km\,s$^{-1}$). Such a merger would likely involve a minor component (if not visible spatially and kinematically) and/or would not significantly broaden the line given the small velocity offset. A similar argument can be made for tidal tails.

Additionally, AGN feedback can produce very rapid outflows of gas which if present, may appear in our composite spectrum. However, AGN-driven winds can often reach thousands of \kms \citep[e.g.][]{bischetti19}, whereas our stacks only show excess emission at velocities of a few hundred \kms. Additionally, the majority of the sample appear to show \nii/\Halpha and \oiii/\Hbeta ratios that place them within the star-formation section of the classical BPT diagram (Faisst et al.\ in prep.).

%
%
\subsection{Outflow velocity}
\label{sec:outflow_velocity}

If we interpret the weak detection of a broad component in the composite spectrum of highly star-forming regions as outflowing gas driven by stellar feedback, we can estimate some basic properties of the potential outflow. Following \cite{lutz20} we measure the outflow velocity as:

\begin{equation}
    v_{\rm out} = |\Delta v| + 0.5{\rm FW}_{10\%},
    \label{eq:v_out}
\end{equation}

\begin{figure}
    \includegraphics[width=\linewidth]{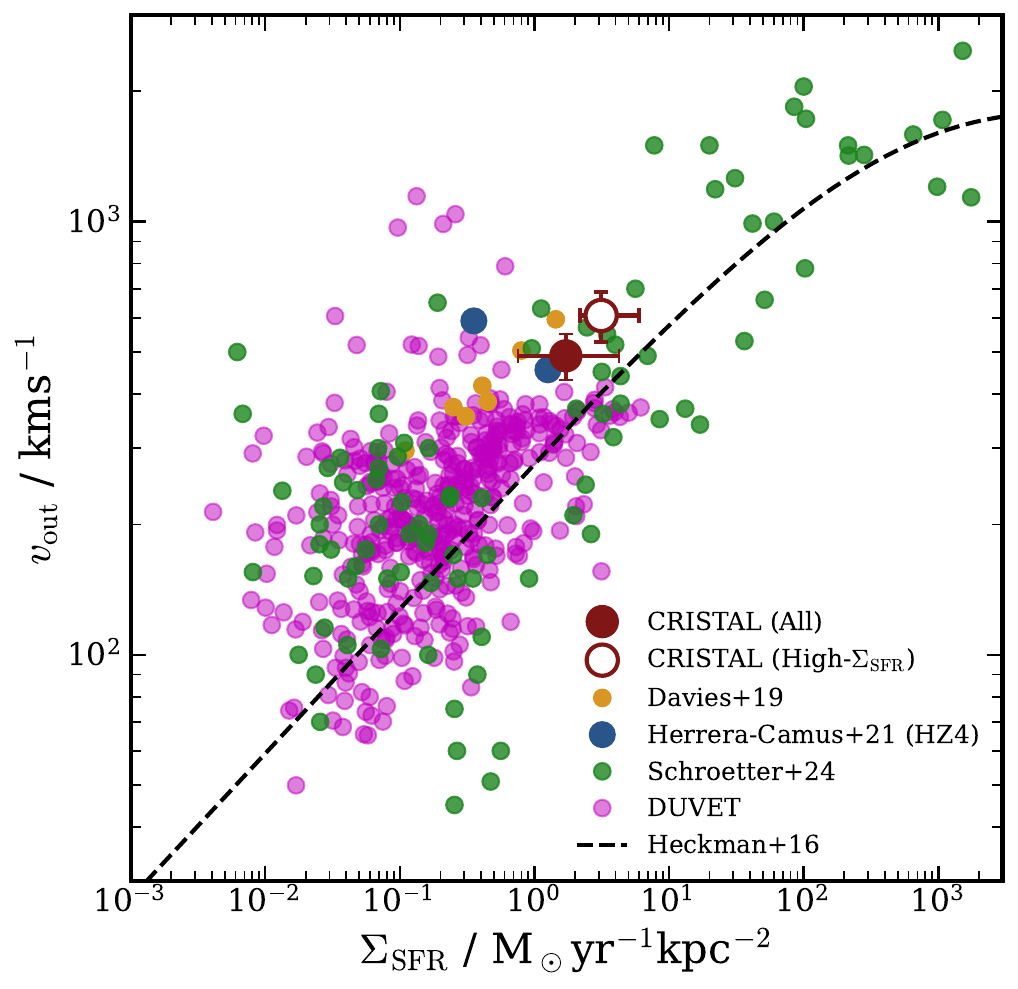}
    \caption{Outflow velocity $v_{\rm out}$ plotted as a function of star-formation rate surface density \sigsfr for the two CRISTAL composites displaying broad emission, along with $z$\,$\sim$\,2--2.6 star-forming galaxies from \cite{davies19} and \cite{schroetter24}. We show the best-fit lines derived by \cite{heckman16} and \cite{chu24} (from the DUVET sample).
    }
    \label{fig:v_out_sig_sfr}
\end{figure}

\noindent where $\Delta v$ is the velocity offset of the broad component and FW$_{10\%}$ is the full-width at 10$\%$ of the peak flux density. For the method 4 stacks of the full sample we find \vout\,$=$\,490\,$\pm$\,59\,km\,s$^{-1}$ and for the high-\sigsfr sample we find \vout\,$=$\,607\,$\pm$\,80\,km\,s$^{-1}$. This is expected if sources with higher star-formation rates drive more powerful outflows, although the two values are consistent within the 1-$\sigma$ uncertainties.

As noted in \S\ref{sec:outflows?}, our sample selection results in disks that are inclined somewhere roughly in the center of edge-on and face-on. If the apparent outflows are perpendicular to the disk then some component on the outflow velocity would not be in the line-of-sight and therefore not detectable in our data. In this case, an inclination correction would be required. For the CRISTAL disks, Lee \etal (in prep.) measured a median inclination of 54$^\circ$, implying a median velocity correction of 1/sin$^2$(54$^\circ$) = 1.24. Applying this would instead yield outflow velocities of \vout\,$=$\,608\,$\pm$\,73\,km\,s$^{-1}$ for the full sample and \vout\,$=$\,753\,$\pm$\,99\,km\,s$^{-1}$ for the high-\sigsfr sample.

Fig.~\ref{fig:v_out_sig_sfr} presents the outflow velocity of the two composites as a function of the star-formation rate surface density, \sigsfr. \sigsfr is calculated as the median value across all pixels included in the stack, with the 16th and 84th percentiles indicated by the error bars. Also included are SINFONI \Halpha measurements by \cite{davies19} for $z$\,$\sim$\,2--2.6 star-forming galaxies, and a literature compilation of $z$\,$<$\,1.5 star-forming galaxies from \cite{schroetter24} (mostly measured from absorption lines of background QSOs). The CRISTAL composites match well with the literature data, and sit close to the measured trend of increasing outflow velocity with star-formation rate surface density measured by both \cite{heckman16} and \cite{chu24}. Although this is not a highly constraining result, it is useful to see that, if interpreted as outflows, our composite properties are comparable to similar samples in the literature.

%
%
\subsection{Mass outflow rate}
\label{sec:mass_outflow_rate}

\begin{figure}
\includegraphics[width=\linewidth]{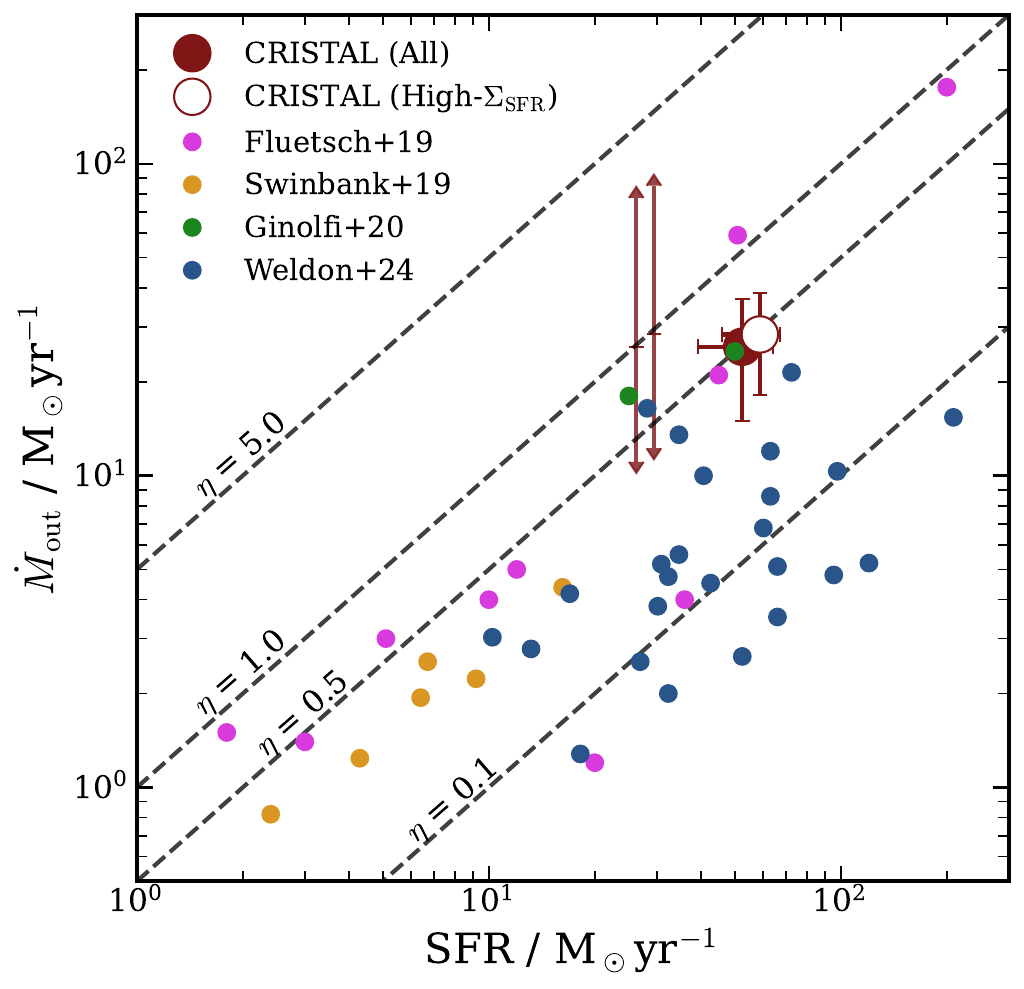}
\caption{Mass outflow rate \moutdot versus star-formation rate for both CRISTAL composite spectra that show evidence for broad emission, along with samples from \cite{swinbank19}, \citetalias{ginolfi20} and \cite{weldon24}. The vertical arrows show how much the two CRISTAL points would move up if we were to adopt a correction factor of 3 to estimate the total mass outflow rate, and how much they would move down if we were to compute the mass outflow rate from only emission in the wings as opposed to the entire broad component (see \S\ref{sec:mass_outflow_rate}).
}
\label{fig:m_out_dot_sfr}
\end{figure}

\begin{figure}
\includegraphics[width=\linewidth]{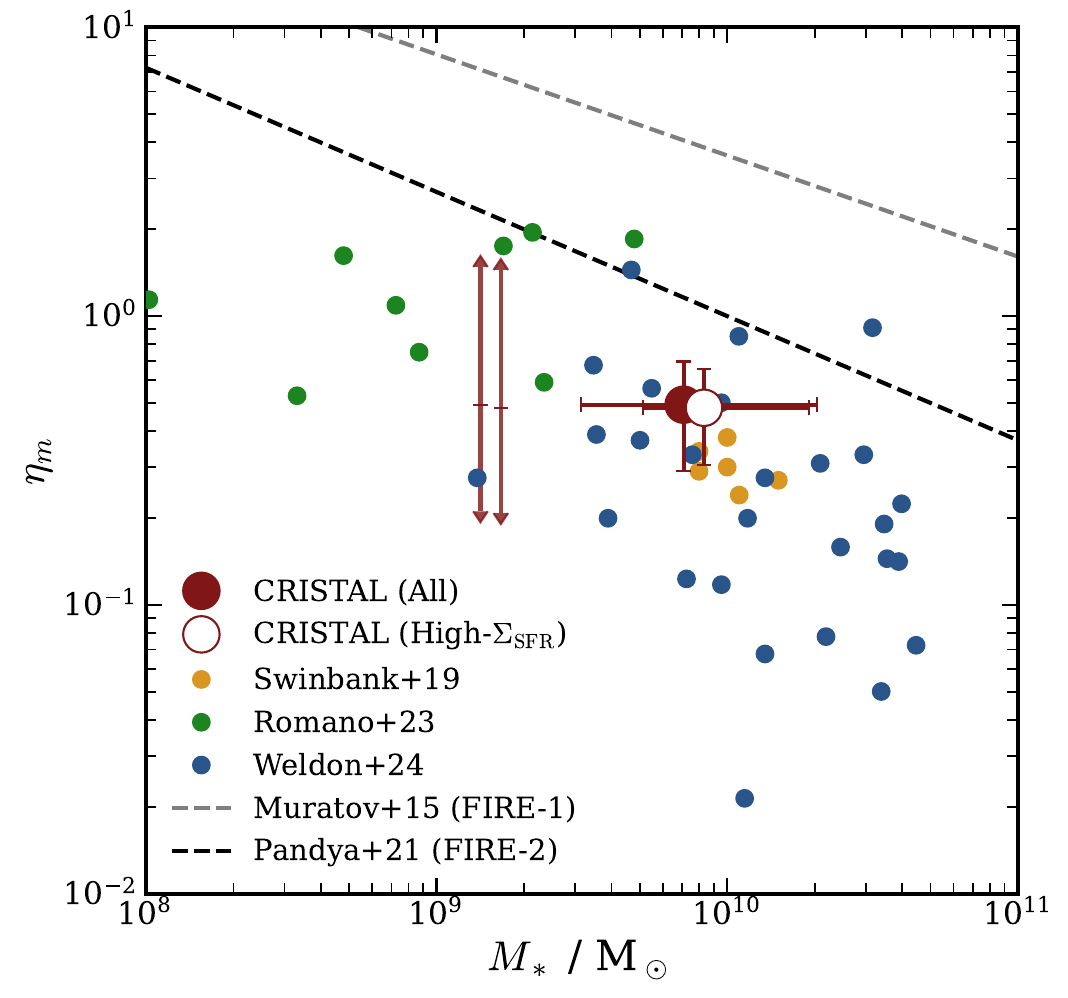}
\caption{Mass loading factor $\eta_m$ plotted as a function of stellar mass. For CRISTAL we show estimates for both the full-sample composite and the high-\sigsfr subsample. The arrows are the same as in Fig.~\ref{fig:m_out_dot_sfr}. For comparison with lower-redshift sources we include results from \protect\cite{swinbank19}, \protect\cite{concas22}, \protect\cite{romano23} and \protect\cite{weldon24} (most of which were inferred from \Halpha data). Predictions from both FIRE-1 and FIRE-2 suggest higher $\eta_m$ than implied by CRISTAL, but we are largely in agreement with the observational data shown.
}
\label{fig:mass_load_mstar}
\end{figure}

Following \citetalias{ginolfi20} we estimate the atomic gas mass of the outflowing region using the following relation from \cite{hailey-dunsheath10}:

\begin{equation}
    \begin{split}
            \dfrac{M_{\rm out}^{\rm atom}}{{\rm M}_\odot} &= 0.7\left(\dfrac{0.7L_{\rm [CII]}}{{\rm L}_\odot}\right) \times \left(\dfrac{1.4\times10^{-4}}{X_{C^+}}\right) \times \\ & \dfrac{1+2e^{-91{\rm K}/T}+n_{\rm crit}/n}{2e^{-91{\rm K}/T}},
    \end{split}
    \label{eq:outflow_mass}
\end{equation}

\noindent where the \cii luminosity $L_{\rm [CII]}$ is derived from the integral of the outflowing regions. We assume a C$^+$ abundance per hydrogen atom of $X_{C^+}$\,$=$\,1.4\,$\times$\,10$^{-4}$, a gas density $n$ equal to the critical density $n_{\rm crit}$\,$=$\,3\,$\times$\,10$^3$\,cm$^{-2}$, and a gas temperature of 100\,K \citep{ginolfi20}. To derive the mass of the outflowing material, we consider the entire broad component of the fit for consistency with \citetalias{ginolfi20}. For an alternative measurement, we use Eq.~\ref{eq:outflow_mass} only including velocity channels where the broad component accounts for at least half of the total flux density, following \cite{lutz20}. This latter approach is more conservative as it aims to exclude flux density contributions from the main component of the galaxy. In Figs.~\ref{fig:m_out_dot_sfr} and \ref{fig:mass_load_mstar}, the downward-pointing red arrows indicate the extent to which this approach reduces the measured values.

We then compute the mass outflow rate as \citep{gallerani18}:
\begin{equation}
    \dot{M}_{\rm out} = \dfrac{v_{\rm out}M_{\rm out}}{R_{\rm out}},
    \label{eq:mass_outflow_rate}
\end{equation}
where $M_{\rm out}$ is derived using Eq.~\ref{eq:outflow_mass}. We adopt $R_{\rm out}$\,$=$\,6\,kpc, following \citetalias{ginolfi20}.

The resulting mass outflow rates are \moutdot\,$=$\,26\,$\pm$\,11\,M$_\odot$\,yr$^{-1}$ for the full sample and \moutdot\,$=$\,28\,$\pm$\,10\,M$_\odot$\,yr$^{-1}$ for the low-\sigsfr sample, although it is important to note that these values assume that only the atomic phase contributes to the outflows. To account for the other ISM phases, we follow recent work by \cite{fluetsch19}, applying a factor of 3 to the atomic outflow rates, yielding total outflow rates of \moutdot\,$=$\,78\,$\pm$\,33\,M$_\odot$\,yr$^{-1}$ for the full sample and \moutdot\,$=$\,85\,$\pm$\,31\,M$_\odot$\,yr$^{-1}$ for the low-\sigsfr sample. We stress that such a correction is based on CO observations of local galaxies, over half of which are Seyferts and LINERs. and therefore may not be applicable to our $z$\,$\sim$\,5 sample. Further, an analysis of multi-phase outflows in M82 by \cite{levy23} found that the corresponding factor would only be $\sim$\,2 \citep[see also work by][]{leroy15,martini18}. In Figs.~\ref{fig:m_out_dot_sfr} and \ref{fig:mass_load_mstar}, the upward-pointing red arrows indicate the extent to which applying the factor of 3 increases the values.

Our mass outflow rate measurements are shown as a function of star-formation rate in Fig.~\ref{fig:m_out_dot_sfr}. Data from the CRISTAL survey are plotted as filled red circles for the entire sample and open red circles for regions with high star-formation rates. Additional outflow measurements from \cite{fluetsch19} and \citetalias{ginolfi20} are shown, along with ionized outflow measurements from \cite{swinbank19} and \cite{weldon24} (at slightly lower redshifts of $z$\,$\sim$\,1--4), although we note that the latter trace a hotter phase of outflowing material, coming from regions more directly impacted by energetic feedback processes like supernovae and stellar winds. This is in contrast with our \cii data, which traces colder outflows. \moutdot and SFR are generally positively correlated, and indeed our sample shows higher outflow rates than lower-SFR literature samples. The high-\sigsfr subset of CRISTAL also shows a marginally higher \moutdot than the full sample, although the two composites are consistent within their uncertainties. Further, our measured \moutdot are consistent with those measured by \citetalias{ginolfi20}. Compared to the other samples, particularly \cite{weldon24}, our composites imply higher mass-loading factors than star-forming galaxies at lower redshifts, if indeed the broad emission we detect is driven by outflows. We discuss this further in the next section.

%
%
\subsection{Mass-loading factor}
\label{sec:mass_loading_factor}

The efficiency of a star-formation driven outflow is typically quantified by the mass-loading factor $\eta_m$\,$=$\,$\dot{M}_{\rm out}/$SFR, that can be intuitively understood as the amount of mass removed per stellar mass newly formed.

For the integrated stack presented in Fig.~\ref{fig:stack_all_2sig} we adopt the median SFR of the sample, 52\,M$_\odot$\,yr$^{-1}$. This gives a mass-loading factor of $\eta_m$\,$=$\,0.49\,$\pm$\,0.20 indicating the mass is lost at a slower rate than new stars are formed. When conservatively multiplying this number by a factor of 3 to account for multi-phase outflows \citep{fluetsch19} this would give $\eta_m$\,$=$\,1.48\,$\pm$\,0.61, consistent with scenarios ranging from no significant overall mass loss to moderate mass loss with time. Given that the \cite{fluetsch19} calibration factor is conservative, we suggest that the former scenario is more likely. For the higher-\sigsfr sample, we find a similar mass-loading factor of $\eta_m$\,$=$\,0.48\,$\pm$\,0.18, implying comparable feedback efficiency between the two samples.

In Fig.~\ref{fig:mass_load_mstar} we show the relationship between stellar mass and the mass-loading factor. Data from the CRISTAL survey are plotted as filled red circles for the entire sample and open red circles for regions with higher star-formation rates, both exhibiting moderate uncertainties around $\eta_m$\,$=$\,0.5. Additional data points from \cite{swinbank19}, \cite{concas22} and \cite{weldon24} are shown, although we again note that these additional data are derived from ionized gas, which traces a hotter phase of the interstellar medium.

For a closer comparison with our data we include results from \cite{romano23} in Fig.~\ref{fig:mass_load_mstar}, which were measured from observations of the \cii emission line in local dwarf galaxies. These systems have stellar masses that are around an order of magnitude lower than the CRISTAL sample, but they generally have comparable mass-loading factors to CRISTAL, and both datasets lie close to the \cite{weldon24} relation. This alignment implies that the feedback efficiency, as represented by $\eta_m$, scales with stellar mass in a consistent manner across cosmic time for the two galaxy types, even though the two samples differ significantly in redshift and galaxy properties.

Theoretical predictions from the FIRE-1 \citep{muratov15} and FIRE-2 \citep{pandya21} simulations are indicated by the dashed gray and black lines, both of which predict a decrease in the mass loading factor as stellar mass increases. This can be interpreted to mean that stellar feedback is more efficient at driving gas outflows in low-mass systems due to their shallower gravitational potential wells. The predictions from FIRE-2 in particular match our full-sample composite reasonably well, but only when adjusted to account for contributions from other phases of outflowing gas.

%
%
\subsection{Contribution of outflows to observed extended \cii emission}

A recent development in the study of high-redshift galaxy evolution is the discovery of apparent extended \cii halos on scales of $\sim$\,10\,kpc. By stacking ALMA data from 18 galaxies at $z$\,$\sim$\,5--7, \cite{fujimoto19} provided the first evidence for this phenomenon, along with several potential driving mechanisms. Currently there is very little consensus on which mechanism is most important, but perhaps the most discussed is outflows removing gas to large radii from the galaxy without it escaping the halo entirely. \cite{pizzati20,pizzati23} developed a semi-analytic model in which supernova-driven outflows can reproduce the \cii halos. In their stacking analysis \citetalias{ginolfi20} used the observed detection of broad emission to suggest that the baryon cycle is already in place in the CGM of normal star-forming galaxies at $z$\,$\sim$\,5.

However, utilizing the CRISTAL sample, \cite{ikeda24} found that the \cii line emission is well-described by a single extended disk component, implying that the aforementioned halos are not as significant as previously thought. Further, they found no correlation between the ratio of $R_{\rm e,[CII]}/R_{\rm e,UV}$ and SFR or $\Sigma_{\rm SFR,IR}$, which they interpreted to suggest that \cii outflows are not the cause of extended emission in the sample. This is clearly consistent with the lack of strong outflows in our full-sample stack (when excluding CRISTAL-02).

%
%
\section{Conclusions}
\label{sec:conclusions}

We have carried out a stacking analysis of ALMA Band 7 data from the CRISTAL survey of $z$\,$\sim$\,5 main-sequence star-forming galaxies. The data cover the forbidden line from singly ionized carbon \cii (rest-frame 158\,$\mu$m), which we use to search for outflow signatures in the form of broad wings. Our main conclusions are as follows:
\begin{itemize}
    \item We adopt a conservative stacking approach, leveraging kinematic analysis of the higher-resolution CRISTAL data (compared to ALPINE) to remove 22 systems which are potentially interacting/merging, leaving a final sample size of 15. When included in a stack, these sources can easily mimic broad wings. Given that CRISTAL-02 is already known to drive a prominent outflow, we perform our stacking analysis both with and without this source.
    \item Given the different redshifts, masses and luminosities of the CRISTAL sources, we test several methods for aggregating the emission line profiles when stacking, including standardizing all linewidths. We find that the chosen method can meaningfully affect the conclusion on whether or not outflows are present, and therefore extreme care must be taken. We prefer to normalize both the linewidths and peak flux densities to a constant value, to ensure that we are probing only the average {\it shape} of the emission line profiles.
    \item After excluding potential mergers from the sample and combining the integrated spectra, the resultant composite shows weak evidence for outflows with \deltabic\,$=$\,17. Evidence with this significance is {\it only} present when normalizing the linewidths and peak flux densities -- all other stacking methods show little-to-no evidence of broad emission in the composites. We measure an outflow velocity (not corrected for inclination) of \vout\,$=$\,490\,$\pm$\,59\,km\,s$^{-1}$, which is consistent with literature measurements for galaxies with similar \sigsfr. The result is also almost entirely driven by CRISTAL-02, without which the significance drops to \deltabic\,$=$\,3.1.
    \item We compare our stacking results with those obtained from the ALPINE survey by \citetalias{ginolfi20}, aiming to identify any systematic differences related to data quality. Although the CRISTAL data offer higher resolution and therefore allow for more effective identification and removal of mergers, our stack reveals less significant broad emission than that reported by \citetalias{ginolfi20}. Moreover, when we inject RMS noise comparable to the CRISTAL data into their composite signal, the resulting stack also shows less significant residuals than those presented in \citetalias{ginolfi20}. We suggest that sample size, selection and methodology all contribute to these differences.
    \item We find a mass outflow rate of \moutdot\,$=$\,26\,$\pm$\,11\,M$_\odot$\,yr$^{-1}$ and mass loading factor of $\eta$\,$=$\,0.49\,$\pm$\,0.20, indicating feedback that, while relatively weak and unlikely to quench star formation, could still regulate the system over long timescales by moderating the gas supply available for star formation.
    \item We exploit the high resolution of the CRISTAL data to generate \cii luminosity maps, which we use as a proxy for star-formation maps. By defining a threshold in star-formation rate, we stack the individual {\it spatial pixels} from the sample which are most highly star forming. When only considering the \cii-brightest regions of the sample (again excluding mergers) which should broadly represent the most highly star forming, the resultant composite spectrum also shows moderate evidence for outflows, \deltabic\,$=$\,9, with \vout\,$=$\,607\,$\pm$\,80\,km\,s$^{-1}$, \moutdot\,$=$\,28\,$\pm$\,10\,M$_\odot$\,yr$^{-1}$, $\eta$\,$=$\,0.48\,$\pm$\,0.18. This could suggest that the potential outflows are driven by star formation.
\end{itemize}
Few sources in the CRISTAL sample individually show evidence for outflows. The analysis presented here indicates that any further evidence would need to be revealed with deeper [C{\sc ii}] observations and/or larger stacking samples. JWST/NIRSpec data covering \Halpha, \nii, \Hbeta and \oiii has recently become available which will push forward our picture of the CRISTAL sample, namely allowing the ruling out of any substantial AGN activity using line ratio diagnostics, the measurement of metallicities, and searching for outflows in the ionized phase.

\begin{acknowledgments}
We would like to thank Michele Ginolfi for useful discussions regarding the ALPINE stacking analysis.
M.A. acknowledges support from ANID Basal Project FB210003 and and ANID MILENIO NCN2024\_112.
T.D.S. acknowledges the research project was supported by the Hellenic Foundation for Research and Innovation (HFRI) under the "2nd Call for HFRI Research Projects to support Faculty Members \& Researchers" (Project Number: 03382).
R.L.D. is supported by the Australian Research Council through the Discovery Early Career Researcher Award (DECRA) Fellowship DE240100136 funded by the Australian Government.
A.F. acknowledges support from the ERC Advanced Grant INTERSTELLAR H2020/740120. Partial support from the Carl Friedrich von Siemens-Forschungspreis der Alexander von Humboldt-Stiftung Research Award is kindly acknowledged.
J.G-L., acknowledges support from ANID BASAL project FB210003 and Programa de Inserción Académica 2024 Vicerrectoría Académica y Prorrectoría Pontificia Universidad Católica de Chile.
R.I. is supported by Grants-in-Aid for Japan Society for the Promotion of Science (JSPS) Fellows (KAKENHI Grant Number 23KJ1006).
K.K. acknowledges support from the Knut and Alice Wallenberg Foundation (KAW 2017.0292).
I.D.L. acknowledges funding from the European Research Council (ERC) under the European Union's Horizon 2020 research and innovation program DustOrigin (ERC-2019- StG-851622) and from the Flemish Fund for Scientific Research (FWO-Vlaanderen) through the research project G0A1523N.
M.R. acknowledges support from project PID2023-150178NB-I00 and PID2020-114414GB-I00, financed by MCIN/AEI/10.13039/501100011033.
M.S. was financially supported by Becas-ANID scholarship \#21221511, and also acknowledges support from ANID BASAL project FB210003.
K.T. acknowledges support from JSPS KAKENHI grant No. 23K03466.
V.V. acknowledges support from the ALMA-ANID Postdoctoral Fellowship under the award ASTRO21-0062.
\end{acknowledgments}

\facilities{ALMA}
\software{{\tt astropy} \citep{astropy18}, CASA \citep{casa}}

\appendix

%
%
\section{Inner and outer galaxy stacks}
\label{sec:inner_outer_stacks}

\begin{figure*}
    \includegraphics[width=\linewidth]{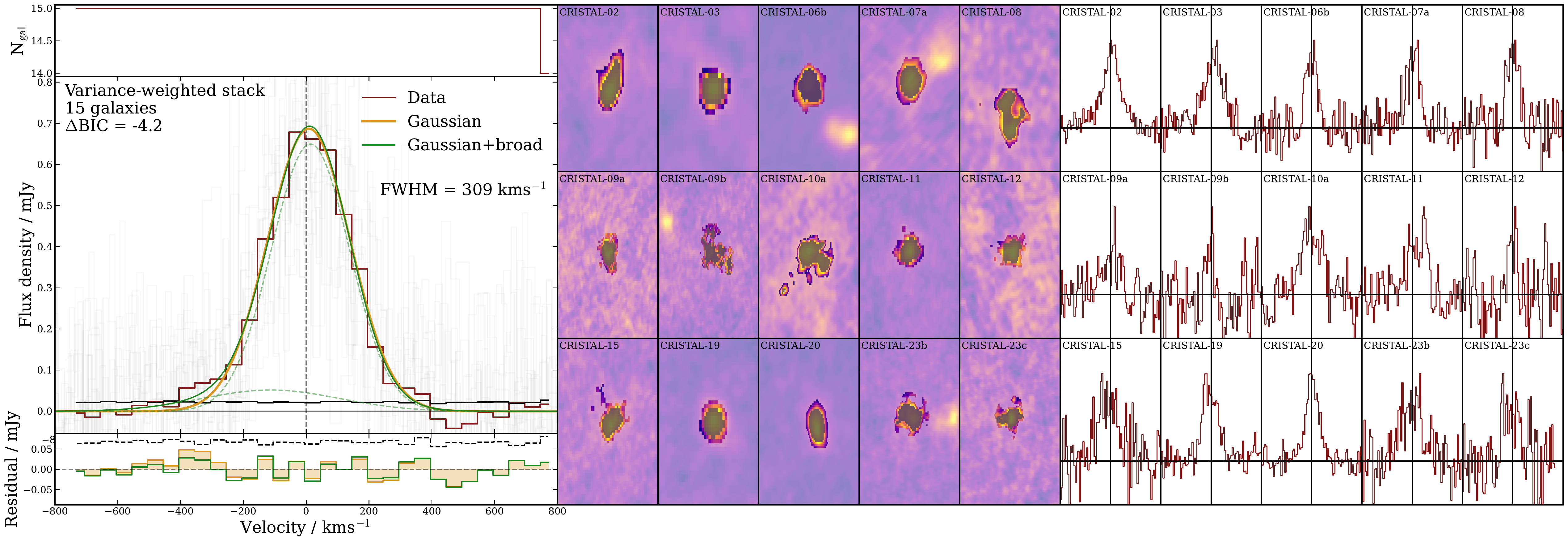}
    \caption{The same as Fig.~\ref{fig:stack_all_2sig}, but only including the outer regions of each galaxy, selected as \sigsfr\,$<$\,$\Sigma_{\rm{SFR,med}}$. The composite is well-modeled by a single-Gaussian function.
    }
    \label{fig:stack_outer}
\end{figure*}

\begin{figure*}
    \includegraphics[width=\linewidth]{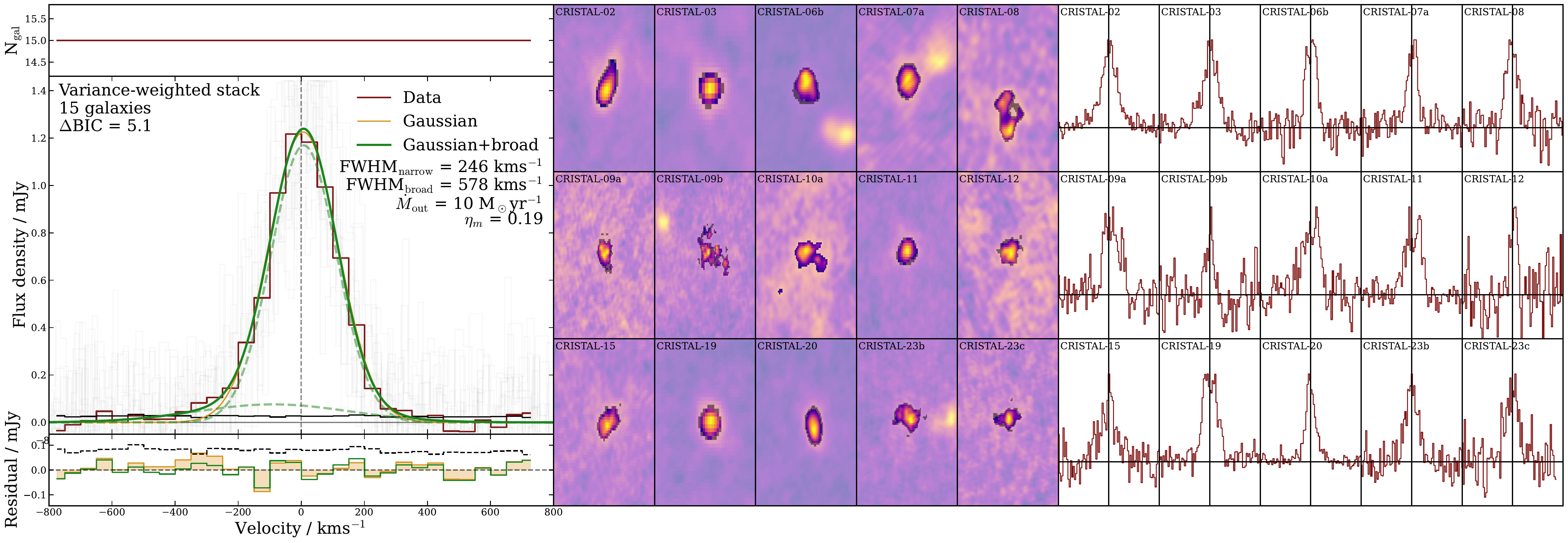}
    \caption{The same as Fig.~\ref{fig:stack_all_2sig}, but only including the inner regions of each galaxy, selected as \sigsfr\,$>$\,$\Sigma_{\rm{SFR,med}}$. The composite is marginally better-modeled when including a broad component.
    }
    \label{fig:stack_inner}
\end{figure*}

\bibliography{bibliography}{}
\bibliographystyle{aasjournal}

\end{document}